\pdfoutput=1
\documentclass[11pt,twoside,a4paper,cmspaper,final,collab]{cms-tdr}
\def\svnVersion{fff7557}\def\svnDate{2025/12/20}\def\cmsCernNoTag{CERN-EP-2025-029}\def\cmsCernDate{\today}\def\cmsMessage{Published in the Journal of High Energy Physics as \href{http://dx.doi.org/10.1007/JHEP12(2025)065}{\doi{10.1007/JHEP12(2025)065}.}}
\begin{document}\cmsNoteHeader{TOP-22-011}

\newlength\cmsTabSkip\setlength{\cmsTabSkip}{2ex}
\providecommand{\cmsTable}[1]{\resizebox{\textwidth}{!}{#1}}

\newcommand{\PAQqpr}{{\HepAntiParticle{\PQq}{}{\prime}}\xspace}
\newcommand{\PQj}{{\HepParticle{j}{}{}}\xspace}
\newcommand{\PQjpr}{{\HepParticle{j}{}{\prime}}\xspace}

\newcommand{\tW}{\ensuremath{\PQt\PW}\xspace}
\newcommand{\tTomutauq}{\ensuremath{\PQt\to\PGm\PGt\PQq}\xspace}
\newcommand{\tTomutauu}{\ensuremath{\PQt\to\PGm\PGt\PQu}\xspace}
\newcommand{\tTomutauc}{\ensuremath{\PQt\to\PGm\PGt\PQc}\xspace}
\newcommand{\BR}{\ensuremath{\mathcal{B}}\xspace}
\newcommand{\STCLFV}{\ensuremath{\text{ST CLFV}}\xspace}
\newcommand{\TTCLFV}{\ensuremath{\text{TT CLFV}}\xspace}
\newcommand{\muR}{\ensuremath{\mu_{\mathrm{R}}}\xspace}
\newcommand{\muF}{\ensuremath{\mu_{\mathrm{F}}}\xspace}
\newcommand{\pp}{\ensuremath{\Pp\Pp}\xspace}
\newcommand{\mtop}{\ensuremath{m_{\PQt}}\xspace}
\newcommand{\mW}{\ensuremath{m_{\PW}}\xspace}
\newcommand{\jj}{\ensuremath{\PQj\PQjpr}\xspace}
\newcommand{\mjj}{\ensuremath{m_{\jj}}\xspace}
\newcommand{\mbjj}{\ensuremath{m_{\PQb\jj}}\xspace}
\newcommand{\chisq}{\ensuremath{\chi^2}\xspace}
\newcommand{\tcmutau}{\ensuremath{\PQt\PQc\PGm\PGt}\xspace}
\newcommand{\tumutau}{\ensuremath{\PQt\PQu\PGm\PGt}\xspace}
\newcommand{\tqmutau}{\ensuremath{\PQt\PQq\PGm{}\PGt}\xspace}
\newcommand{\mutauh}{\ensuremath{\PGm\tauh}\xspace}
\newcommand{\CLFVWC}[2]{\ensuremath{C^{\mkern1mu\text{#1}}_{#2}}\xspace}

\cmsNoteHeader{TOP-22-011}

\title{Search for charged-lepton flavour violation in top quark interactions with an up-type quark, a muon, and a \texorpdfstring{\PGt}{tau} lepton in proton-proton collisions at \texorpdfstring{$\sqrt{s}=13\TeV$}{sqrt(s) = 13 TeV}}

\date{\today}

\abstract{A search for charged-lepton flavour violation (CLFV) in top quark (\PQt) production and decay is presented. The search uses proton-proton collision data corresponding to 138\fbinv collected with the CMS experiment at $\sqrt{s}=13\TeV$. The signal consists of the production of a single top quark via a CLFV interaction or top quark pair production followed by a CLFV decay. The analysis selects events containing a hadronically decaying \PGt lepton and a muon of opposite electric charge, as well as at least three jets, one of which is identified as originating from the fragmentation of a bottom quark. Machine learning classification techniques are used to distinguish signal from standard model background events. The results of this search are consistent with the standard model expectations. The upper limits at 95\% confidence level on the branching fraction \BR for CLFV top quark decays to a muon, a \PGt lepton, and an up or a charm quark are set at $\BR(\tTomutauu)<(0.04$, 0.08, and $0.12)\times10^{-6}$, and $\BR(\tTomutauc)<(0.81$, 1.71, and $2.05)\times10^{-6}$ for scalar, vector, and tensor-like operators, respectively.}

\hypersetup{%
pdfauthor={CMS Collaboration},%
pdftitle={Search for charged-lepton flavour violation in top quark interactions with an up-type quark, a muon, and a tau lepton in proton-proton collisions at sqrt(s) = 13 TeV},%
pdfsubject={CMS},%
pdfkeywords={CMS, Top, LFV, Tau}}

\maketitle
\section{Introduction}

The standard model (SM) of particle physics successfully describes the properties of all
elementary particles and their interactions with a small number of free parameters, but it is known to be incomplete.
Extensions of the SM aim to describe several phenomena such as dark matter~\cite{Gaitskell:2004gd, Trimble:1987ee, Porter:2011nv, ref:DMevi} and the baryon asymmetry of the universe~\cite{Sakharov:1967dj}.

The observation of neutrino oscillations~\cite{neutosc} confirms a non-zero neutrino mass
and neutral-lepton flavour violation.
The existence of neutrino mass terms implies charged-lepton flavour violation (CLFV) in the SM at the loop level.
An example is the CLFV branching fraction \BR of a muon decay into an electron and a photon,
which is of the order of $10^{-55}$ in the SM, whereas extensions of the SM predict branching fractions
as high as $10^{-4}$~\cite{Calibbi:2017uvl}.
The observation of CLFV in the top quark sector at the LHC would be a decisive signature of physics beyond the SM.

Recent combinations of measured fractions of \PB meson decays into \PD and \PDst mesons~\cite{HFLAV:2024ctg} involving a \PGt lepton are in tension with the corresponding SM predictions.
This has sparked interest in models that aim to explain these deviations, and at the same time introduce
a new source of lepton flavour violation (LFV) effects.
For example, leptoquark models developed to explain the anomalies in the \PQb quark sector
could generate a four-fermion interaction that would allow the $\PQt\to\PQc\Pell_i\Pell_j$
decay at tree level~\cite{Kim:2018oih}, where \Pell denotes a charged lepton
and the indices $i$ and $j$ denote the different lepton flavours, with $i \ne j$.
Recently, the Belle II Collaboration reported a 1.7 standard deviation difference from the SM predictions of the branching
fraction ratios~\cite{Belle-II:2025yjp}.
Although the new measurement aligns more closely with the SM expectations than the combined measurement, the interest in the search for LFV remains.

The ATLAS Collaboration has performed a search for CLFV in $\PGm\PGt\PQq\PQt$ interactions using
proton-proton (\pp) collision data corresponding to 140\fbinv at $\sqrt{s}=13\TeV$~\cite{ATLAS:2024njy}.
Events with two muons, one hadronically decaying \PGt lepton (\tauh),
and at least one jet identified to originate from the fragmentation of a \PQb quark (\PQb jet) are analysed,
resulting in an upper limit at 95\% confidence level (\CL) of $\BR(\tTomutauq)<8.7\times10^{-7}$,
where \PQq can be either a \PQu or \PQc quark.
The CMS Collaboration has performed searches for CLFV involving $\Pe\PGm\PQt\PQq$ interactions
in final states with two or three leptons~\cite{TOP-19-006, CMS:2023iul}. The upper limits at 95\% \CL
on the corresponding branching fractions range from $10^{-6}$ to $10^{-8}$.

\begin{figure}[!b]
\centering
\includegraphics[width=0.32\textwidth]{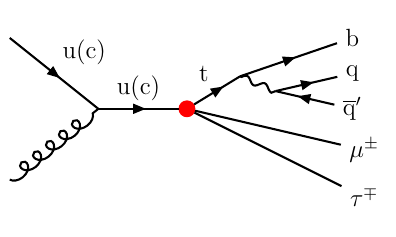}%
\hfil%
\includegraphics[width=0.32\textwidth]{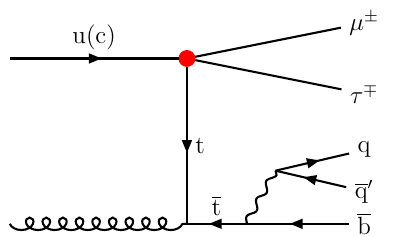}%
\hfil%
\includegraphics[width=0.32\textwidth]{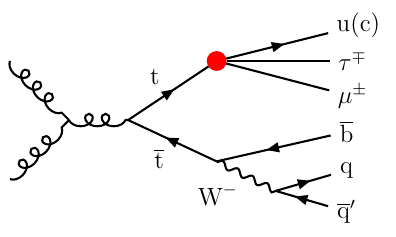}
\caption{%
    Example Feynman diagrams at leading order for the CLFV production of a single top quark
    (left and centre) and top quark pair production followed by a CLFV decay (right).
    Red dots indicate the effective CLFV coupling.
}
\label{fig:Feynman}
\end{figure}

A new search for CLFV involving interactions with top quarks is presented in this article.
We focus on final states with a muon and a \tauh with opposite charge, and multiple jets.
The analysis uses \pp collision data collected with the CMS detector at $\sqrt{s}=13\TeV$,
corresponding to an integrated luminosity of 138\fbinv.
The analysis focuses on CLFV single top quark (\STCLFV) production and top quark pair (\ttbar)
production with a subsequent CLFV decay (\TTCLFV), where the SM top quark decays into jets, as illustrated in Fig.~\ref{fig:Feynman}.
Signal events are modelled using an effective field theory (EFT) framework with dimension-six operators $O_a^{(6)}$ and the corresponding Wilson coefficients $C_a^{(6)}$, where the generic Lagrangian is given by
\begin{equation}
    \mathcal{L}_{\text{eff}}=\mathcal{L}_{\text{SM}} +
    \frac{1}{\Lambda^2} \sum_a C_a^{(6)} O_a^{(6)} + \mathcal{O}\left(\frac{1}{\Lambda^4}\right).
    \label{eq:EFT_Lagrangian}
\end{equation}
The sum runs over the relevant dimension-six operators introducing CLFV effects and the new physics scale $\Lambda$ is set to 1\TeV.
Terms of order $1/\Lambda^4$ and higher are neglected~\cite{Aguilar-Saavedra:2018ksv, Grzadkowski}.
Table~\ref{tab:EFToperators} summarizes the relevant EFT operators considered in this analysis
and their grouping according to the Lorentz structure.
In this analysis, the $O_a^{(6)}$ are defined as linear combinations of the operators given in Table~\ref{tab:EFToperators}.
The analysis considers six independent Wilson coefficients that lead to CLFV couplings:
\CLFVWC{tensor}{\tcmutau}, \CLFVWC{vector}{\tcmutau}, \CLFVWC{scalar}{\tcmutau},
\CLFVWC{tensor}{\tumutau}, \CLFVWC{vector}{\tumutau}, \CLFVWC{scalar}{\tumutau}.
For each coefficient, the subscript denotes the CLFV interaction vertex and
the superscript represents the Lorentz structure of the corresponding operator.

\begin{table}[!tb]
\centering
\topcaption{%
    The EFT operators considered in this analysis and their definition.
    The $\varepsilon$ parameter is a fully asymmetric two-dimensional tensor,
    $\gamma^{\mu}$ are the Dirac matrices, and $\sigma^{\mu\nu}=\frac{i}{2}[\gamma^{\mu},\gamma^{\nu}]$.
    Left-handed doublets of leptons and quarks are denoted by $\Pell_i$ and $\PQq_k$, respectively,
    where the indices $i$ and $k$ denote the lepton and quark flavours. Right-handed lepton and quark singlets
    are denoted by $\Pe_i$ and $\PQu_k$, respectively.
}
\renewcommand{\arraystretch}{1.5}
\begin{tabular}{clcc}
    Structure & Operator & Definition \\ \hline
    Scalar & $O_{\Pell\Pe\PQq\PQu}^{1(ijkl)}$ & $(\overline{\Pell}_{i}\Pe_{j}) \varepsilon (\PAQq_{k}\PQu_{l})$ \\[\cmsTabSkip]
    \multirow{4}{*}{Vector} & $O_{\Pell\PQq}^{1(ijkl)}$ & $(\overline{\Pell}_{i}\gamma^{\mu}\Pell_{j})(\PAQq_{k}\gamma_{\mu}\PQq_{l})$ \\
        & $O_{\Pell\PQu}^{(ijkl)}$ & $(\overline{\Pell}_{i}\gamma^{\mu}\Pell_{j})(\PAQu_{k}\gamma_{\mu}\PQu_{l})$  \\
        & $O_{\Pe\PQq}^{(ijkl)}$ & $(\overline{\Pe}_{i}\gamma^{\mu}\Pe_{j})(\PAQq_{k}\gamma_{\mu}\PQq_{l})$ \\
        & $O_{\Pe\PQu}^{(ijkl)}$ & $(\overline{\Pe}_{i}\gamma^{\mu}\Pe_{j})(\PAQu_{k}\gamma_{\mu}\PQu_{l})$ \\[\cmsTabSkip]
    Tensor & $O_{\Pell\Pe\PQq\PQu}^{3(ijkl)}$ & $(\overline{\Pell}_{i}\sigma^{\mu\nu}\Pe_{j}) \varepsilon (\PAQq_{k}\sigma_{\mu\nu}\PQu_{l})$ \\
\end{tabular}
\label{tab:EFToperators}
\end{table}

This article is organized as follows.
We start with a brief overview of the CMS detector in Section~\ref{sec:cms}.
A detailed description of the signal and background modelling is presented in Section~\ref{sec:sample}.
In Section~\ref{sec:objects}, we describe the objects and the selection requirements, followed
by improvements of the background modelling in Section~\ref{sec:bkgestimation}.
The reconstruction of the top quarks and \PW bosons, as well as the signal discrimination
are explained in Section~\ref{sec:signalextraction}. Systematic uncertainties
are listed in Section~\ref{sec:systematic}.
The results of the search are presented in Section~\ref{sec:results}.
We conclude with a summary in Section~\ref{sec:conclusion}.
Tabulated results are provided in the HEPData record for this analysis~\cite{hepdata}.

\section{The CMS detector}
\label{sec:cms}

The central feature of the CMS apparatus is a superconducting solenoid of 6\unit{m} internal diameter,
providing a magnetic field of 3.8\unit{T}.
Within the solenoid volume are a silicon pixel and strip tracker, a lead tungstate crystal
electromagnetic calorimeter (ECAL), a brass and scintillator hadron calorimeter (HCAL),
each composed of a barrel and two endcap sections.
Forward calorimeters extend the pseudorapidity ($\eta$) coverage provided by the barrel and endcap detectors.

Events of interest are selected using a two-tiered trigger system. The first level,
composed of custom hardware processors, uses information from the calorimeters and muon detectors
to select events at a rate of around 100\unit{kHz} within a fixed latency of about 4\mus~\cite{CMS:2020cmk}.
The second level, known as the high-level trigger, consists of a farm of processors
running a version of the full event reconstruction software optimized for fast processing,
and reduces the event rate to around 1\unit{kHz} before data storage~\cite{CMS:2016ngn}.
A more detailed description of the CMS detector, together with a definition of the coordinate
system used and the relevant kinematic variables, can be found in Ref.~\cite{CMS:2008xjf}.

\section{Data and simulated samples}
\label{sec:sample}

Data were collected with the CMS detector in \pp collisions
in the years 2016 to 2018 at $\sqrt{s}=13\TeV$, corresponding to an integrated luminosity of 138\fbinv.
Data and simulation are analyzed in four distinct data-taking periods, taking into account significant changes in the detector configuration that affected tracking efficiency in 2016.
The data analysed in this search were recorded by triggers requiring
the presence of a single muon.
We use high-level trigger paths with isolated muons~\cite{CMS:2018rym}, where the muon transverse momentum (\pt) thresholds
vary between 24 and 27\GeV, depending on the data-taking period.

Signal CLFV events are generated at leading order (LO) accuracy with \MGvATNLO 2.6.5~\cite{Alwall:2014hca},
separately for single top quark and \ttbar production using the \textsc{SmeftFR} model~\cite{Dedes:2019uzs}.
The \STCLFV cross sections are calculated at LO with \MGvATNLO.
The \TTCLFV signal cross sections are calculated as product of the SM \ttbar cross section of 833.9\unit{pb}
as obtained from the \TOPpp program at next-to-next-to-LO (NNLO) in perturbative quantum
chromodynamics (QCD) including soft-gluon resummation at next-to-next-to-leading-logarithmic
(NNLL) order~\cite{Czakon:2011xx} with the CLFV top quark branching fractions.
These are obtained from the partial widths of the CLFV top quark decay from
Refs.~\cite{Kile:2008rp,AguilarSaavedra:2010zi}, which do not depend on the flavour of the up-type quark (u or c) involved in the CLFV vertex.
The distributions and normalization of the signal samples are predicted using \textsc{SMEFTsim}~\cite{Brivio:2020onw} as done in Ref.~\cite{CMS:2023iul}. 
Table~\ref{tab:totcalc_xsec} summarizes the predicted cross sections
for each channel considered with the Wilson coefficients set to $C_a/\Lambda^2=1\TeV^{-2}$.

\begin{table}[!tb]
\renewcommand{\arraystretch}{1.2}
\centering
\topcaption{%
    Predicted cross sections for CLFV signal processes together with uncertainties from missing higher orders in matrix element calculations, considering operators with different Lorentz structures using $C_a/\Lambda^2=1\TeV^{-2}$.
    The results for \STCLFV are at LO accuracy and the ones
    for \TTCLFV are at NNLO+NNLL accuracy for the \ttbar production
    with LO accuracy for the CLFV decay.
    The cross sections are calculated for the $\PGm\PGt$ flavour combination only, assuming a single quark flavour (either \PQu or \PQc) in the CLFV interaction, and include the branching fraction of the SM top quark in the fully hadronic decay channel.
}
\label{tab:totcalc_xsec}
\renewcommand{\arraystretch}{1.3}
\begin{tabular}{ccc}
Process & Lorentz structure & Cross section [fb] \\ \hline
    \multirow{3}{*}{\STCLFV \tumutau}
    & Scalar & $59.1\,^{+11.1}_{-8.7}$  \\
    & Vector & $276\,^{+50}_{-39}$  \\
    & Tensor & $1272\,^{+220}_{-180}$  \\[\cmsTabSkip]
    \multirow{3}{*}{\STCLFV \tcmutau}
    & Scalar & $3.74\,^{+0.63}_{-0.54}$  \\
    & Vector & $19.5\,^{+3.1}_{-2.6}$   \\
    & Tensor & $96\,^{+14}_{-12}$  \\[\cmsTabSkip]
    \multirow{3}{*}{\TTCLFV \tqmutau}
    & Scalar & $2.69\,^{+0.71}_{-0.53}$  \\
    & Vector & $21.5\,^{+5.7}_{-4.2}$  \\
    & Tensor & $129\,^{+34}_{-25}$  \\
\end{tabular}
\end{table}

The top quark \pt spectrum in \ttbar production simulated at LO and next-to-LO (NLO)
differs from higher-order theoretical predictions~\cite{CMS:TOP-17-014, ATLAS:2016pbv}.
Therefore, the distribution in the top quark \pt in the \TTCLFV events is corrected sequentially
from LO to NLO and to the NNLO QCD+NLO electroweak prediction
utilizing simulated SM \ttbar samples and theory predictions~\cite{Czakon:2017wor,CMS:2021vhb}, respectively.

Background processes are divided into three groups: \ttbar production,
single top quark production, and the other minor backgrounds.
These are referred to as ``\ttbar'', ``single \PQt'', and ``other'', respectively.
The largest background process is SM \ttbar  production in dileptonic decay channels and lepton+jets channels.
It is simulated at NLO precision using \POWHEG v2~\cite{powheg1,powheg2,powheg3}.
The \pt spectrum in the SM \ttbar sample is corrected to match the NNLO predictions,
similar to the \TTCLFV signal samples.
The single \PQt background consists of $t$-channel and $s$-channel processes,
and \PW boson associated production (\tW). These are simulated at NLO accuracy
using \POWHEG v2, \MGvATNLO, and \POWHEG v2~\cite{Re:2010bp}, respectively.
The predicted NNLO cross sections are 134 and 80\unit{pb} for $t$-channel top quark and antiquark production, respectively,
and 6.8\unit{pb} for $s$-channel production~\cite{Campbell:2020fhf}.
The \tW cross section is 79.3\unit{pb}, calculated at NNLO accuracy in QCD with soft-gluon emissions by resuming NNLL terms~\cite{Kidonakis:2021vob}.
Other smaller SM background contributions arise from \ttbar production in fully hadronic decays,
\ttbar production in association with a vector boson ($\ttbar\PZ$ and $\ttbar\PW$)
or a Higgs boson ($\ttbar\PH$), simulated with \MGvATNLO and \POWHEG v2 at NLO accuracy, respectively,
Drell--Yan and \PW boson production, simulated with \MGvATNLO at LO with up to 4 additional
jets using MLM matching~\cite{Alwall:2007fs}, and diboson ($\PW\PW$, $\PW\PZ$, and $\PZ\PZ$)
processes generated with \PYTHIA~8.240~\cite{Sjostrand:2014zea} at LO accuracy.
A top quark mass of 172.5\GeV is used in all simulated samples.
The parton distribution function (PDF) set used for the event generation is NNLO NNPDF~3.1~\cite{NNPDF:2017mvq},
while \PYTHIA with the CP5 tune~\cite{Sirunyan:2019dfx} is used to simulate initial- and final-state radiation,
the parton shower, and the hadronization process.
All simulated samples are interfaced to \GEANTfour~\cite{Agostinelli:2002hh}
for a detailed simulation of the CMS detector response.

{\tolerance=800
Additional \pp interactions within the same or nearby bunch crossings (pileup)
are simulated by generating inelastic \pp events using \PYTHIA
with a total inelastic cross section of 69.2\unit{mb}~\cite{CMS:2018mlc}.
These are superimposed on the hard-scattering events.
The generated number of \pp interactions is corrected such that
the simulated distribution in the number of primary vertices matches that observed in the data.
\par}

\section{Object reconstruction and event selection}
\label{sec:objects}

The particle-flow (PF) algorithm~\cite{CMS:2017yfk} aims to reconstruct and identify
each individual particle in an event, with an optimized combination of information
from the various elements of the CMS detector including charged-particle tracks
from the tracking detector, energy deposits in the HCAL and ECAL, and reconstructed
tracks from the muon chambers. Particles are reconstructed and identified
as either electrons, muons, photons, charged hadrons, or neutral hadrons.
The primary vertex is taken to be the vertex corresponding to the hardest scattering in the event,
evaluated using tracking information alone, as described in Section 9.4.1 of Ref.~\cite{CMS-TDR-15-02}.

Muons are identified as tracks in the central tracker consistent with either a track or
several hits in the muon system, and associated with calorimeter deposits compatible with the muon hypothesis.
The momentum of muons is obtained from the curvature of the corresponding track.
Muons are required to have $\pt>50\GeV$ and $\abs{\eta}<2.4$.
This $\pt$ threshold targets energetic muons that are characteristic of signal events, particularly in \STCLFV, and helps suppress background contributions.
Moreover, muons need to be isolated and pass the ``tight'' identification criteria,
which corresponds to an efficiency of 96\%~\cite{CMS:2018rym}.
The muons passing ``loose'' isolation and identification criteria are
used to veto events with additional muons with $\pt>15\GeV$.

Electrons are identified as a primary charged-particle track and potentially multiple ECAL
energy clusters corresponding to the extrapolation of this track to the ECAL and to
possible bremsstrahlung photons emitted along the way.
The electron momentum is estimated by combining the energy measurement in the ECAL
with the momentum measurement in the tracker.
Only electrons passing the ``veto'' cut-based identification~\cite{CMS:2020uim} with $\pt>15\GeV$ and $\abs{\eta}<2.4$
are considered as these are used to reject events with additional leptons.

Charged hadrons are identified as charged-particle tracks that are neither identified as electrons nor muons.
Finally, neutral hadrons are identified as HCAL energy clusters not linked to any charged-hadron
trajectory, or as a combined ECAL and HCAL energy excess with respect to the expected
charged-hadron energy deposit. The energy of charged hadrons is determined from a combination
of their momentum measured in the tracker and the matching ECAL and HCAL energy deposits,
corrected for the response function of the calorimeters to hadronic showers.
The energy of neutral hadrons is obtained from the corresponding corrected ECAL and HCAL energy deposits.

The PF candidates are clustered into jets using the anti-\kt clustering algorithm~\cite{Cacciari:2008gp}
with a distance parameter of 0.4, implemented in the \FASTJET package~\cite{Cacciari:2011ma,Cacciari:2005hq}.
The jet momentum is determined as the vectorial sum of all particle momenta in the jet,
and is found from simulation to be, on average, within 5--10\% of the true momentum over
the entire \pt spectrum and detector acceptance.
Pileup can contribute additional tracks and calorimetric energy depositions to the jet momentum.
To mitigate this effect, tracks identified as originating from pileup vertices are
discarded and a correction is applied for the remaining contributions.
Jet energy corrections are derived from simulation studies so that the average measured
energy of reconstructed jets becomes identical to that of particle-level jets.
In situ measurements of the momentum balance in dijet, photon+jet, \PZ{}+jet, and multijet
events are used to determine any residual differences between the jet energy scale (JES)
in data and in simulation, and appropriate corrections are made~\cite{CMS:2016lmd}.
The jet energy resolution (JER) amounts typically to 15--20\% at 30\GeV, 10\% at 100\GeV,
and 5\% at 1\TeV~\cite{CMS:2016lmd}. Heavy-flavour jets are identified based on the
\textsc{DeepJet} algorithm~\cite{BTV-16-002,Bols:2020bkb,CMS-DP-2018-058} at the medium working point
defined by a light-quark or gluon jet misidentification rate of 1\%, resulting in a \PQb jet
identification efficiency of 75\%. Only jets with $\pt>40\GeV$ and $\abs{\eta}<2.4$
are considered.

Hadronic \PGt lepton decays are reconstructed from jets, using the hadrons-plus-strips
algorithm \cite{CMS:2018jrd}, which combines one or three tracks with energy deposits in
the calorimeters, to identify the \tauh lepton decay modes and charge.
Neutral pions are reconstructed as strips with dynamic size in $\eta$-$\phi$, with
$\phi$ denoting the azimuthal angle in radians, from reconstructed electrons and photons~\cite{CMS:2020uim}.
The strip size varies as a function of the \pt of the electron or photon candidate.
To distinguish genuine \tauh decays from jets originating from the hadronization of quarks or gluons,
and from electrons or muons, the \textsc{DeepTau} algorithm is used~\cite{CMS:2022prd}.
Information from all individual reconstructed particles near the \tauh axis is combined
with properties of the \tauh candidate and the event.
The \tauh candidates must satisfy kinematic requirements of $\pt>40\GeV$ and $\abs{\eta}<2.3$.
Candidates are selected using the ``very tight'' criterion to discriminate against quark and gluon jets,
and the ``tight'' and ``very loose'' criteria against muons and electrons, respectively.
The chosen identification working points have \tauh average efficiencies of about 50, 98, and 98.8\%
with misidentification probabilities of about 0.4, 2, and 0.02\% for the discrimination against
jets, electrons, and muons, respectively~\cite{CMS:2022prd}.
The rate of a jet to be misidentified as \tauh by the \textsc{DeepTau} algorithm depends on
the \pt and quark flavour of the jet.
The \tauh energy scale corrections are derived from a binned maximum likelihood fit of
the distributions of observables sensitive to the energy shift~\cite{CMS:2018jrd,CMS:2022prd}.

The missing transverse momentum vector \ptvecmiss is computed as the negative vector
\pt sum of all the PF candidates in an event, and its magnitude is denoted as \ptmiss~\cite{Sirunyan:2019kia}.
The \ptvecmiss is modified to account for corrections to the energy scale of the reconstructed jets in the event.

The event selection is designed to isolate final-state particles characteristic of CLFV signal
processes in both single top quark and top quark pair production.
Events are selected via a sequence of requirements on selected objects.
Events must have exactly one tight muon with no additional muons or electrons.
Furthermore, events must contain exactly one \tauh candidate,
separated by $\DR>0.4$ from the muon.
The angular distance \DR is defined as
$\DR=\sqrt{\smash[b]{(\Delta\eta)^2+(\Delta\phi)^2}}$, where $\Delta\eta$ and $\Delta\phi$
are the differences in pseudorapidity and azimuthal angle between the \tauh and
muon candidates, respectively.
Only events containing an oppositely charged muon-\tauh pair are considered.
We remove jets with $\DR<0.4$ from the selected muon and \tauh candidates.
Events must contain at least three jets, where exactly one jet must be a \PQb jet.

The lepton identification efficiencies for the muon and \tauh candidates used in this analysis
are measured using the ``tag-and-probe'' technique~\cite{CMS:2011aa}.
Efficiency differences between data and simulation are corrected for by applying scale factors (SFs)
to the simulation.

\section{Background from jets misidentified as \texorpdfstring{\tauh}{tau}}
\label{sec:bkgestimation}

Despite the discrimination power of the \textsc{DeepTau} algorithm with a \tauh misidentification probability
of less than 1\% for quark and gluon jets, a non-negligible background contribution remains
from events with misidentified \tauh candidates.
Most of these are \ttbar events in the lepton+jets decay channel,
where the correctly identified muon originates from the top quark decay chain $\PQt\to\PW\PQb\to\PGm\PGn\PQb$.
Other events with a misidentified \tauh candidate originate from \PW{}+jet
and single \PQt production. Because of the much smaller muon misidentification rate,
events with a misidentified muon and a genuine \tauh candidate are negligible compared
to events with a misidentified \tauh candidate.
The rate of misidentified \tauh candidates is not well modelled by simulation,
therefore an approach based on control regions~(CRs) in data is used to determine the contribution of events
with a misidentified \tauh candidate in this analysis.

The ``ABCD method'', as introduced in Ref.~\cite{CDF:1990kbl}, uses additional
event categories to calculate scaling factors for simulated events to correct
for a mismodelling in the simulation.
The signal region (SR) D contains events that pass all selections of Section~\ref{sec:objects},
with muon and \tauh candidates having opposite charges.
The CR B is defined with the muon and \tauh candidates having the same charge.
Additional CRs are constructed by requiring events to have a \tauh candidate passing
the ``loose'', but failing the ``very tight'' requirement of the \tauh discrimination against jets.
These events enter CR A if the muon and \tauh candidates have the same charge, and CR C
if they have opposite charges.
The same requirements on the jet and \PQb jet multiplicities are applied to all regions.
The number of events with a misidentified \tauh candidate in SR D, $N^{\text{D}}_{\text{mis-ID}}$,
is calculated from data in CRs A, B, and C, as
\begin{equation}
   N^{\text{D}}_{\text{mis-ID}} = N^{\text{C}} \frac{N^{\text{B}}}{N^{\text{A}}}.
   \label{eq:abcd1}
\end{equation}
In each region $i$, the number of events $N^i$ entering Eq.~\eqref{eq:abcd1}
is the number of observed events in data $N^i_{\text{obs}}$ corrected by the number of
events with a genuine \tauh lepton $N^i_{\text{genuine}}$,
\begin{equation}
  N^i = N^i_{\text{obs}} - N^i_{\text{genuine}}.
\end{equation}
The value of $N^i_{\text{genuine}}$ is determined using generator-level information in simulated events,
which contains the properties of particles as defined by the event generator before the detector simulation.
The ABCD method corrects the normalization of simulated background processes with
misidentified \tauh candidates in the SR D under the assumption that
the ratio between the number of jets misidentified as \tauh passing the ``very tight'' and ``loose''
selection criteria is the same for same-sign and opposite-sign events.
The SF applied to simulated background events is calculated as
$N^{\mathrm{D}}_{\text{mis-ID}}/(N^{\mathrm{D}}_{\text{tot. bkg.}} - N^{\mathrm{D}}_{\text{genuine}})$,
where $N^{\mathrm{D}}_{\text{tot. bkg.}}$ is the number of all simulated background events in SR D.
This SF is calculated and applied separately for each data-taking period
and \tauh decay mode, considering the number of charged and neutral hadrons in the \tauh decay products.

\begin{figure}[!tbp]
\centering
\includegraphics[width=0.49\textwidth]{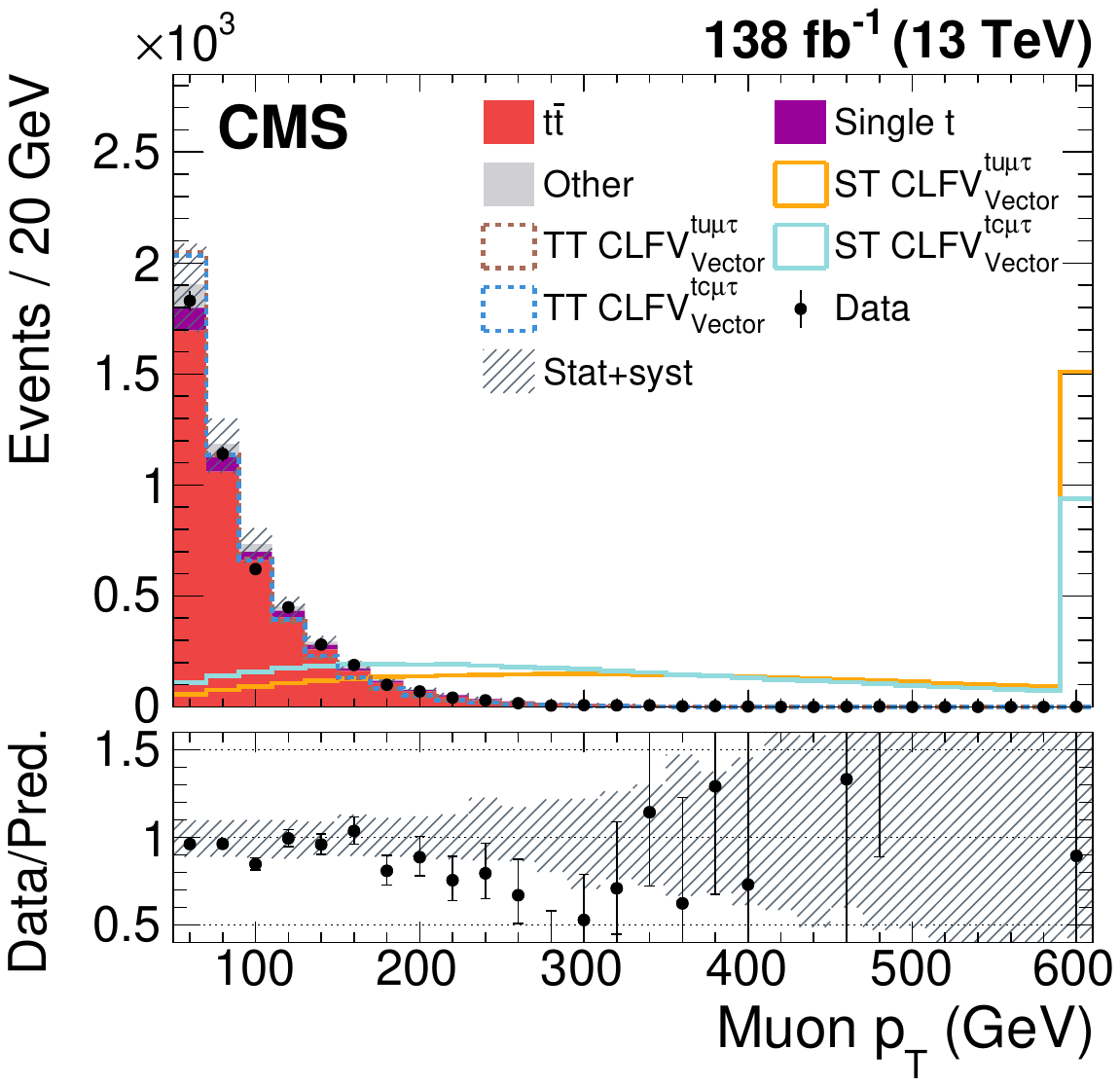}%
\hfill%
\includegraphics[width=0.49\textwidth]{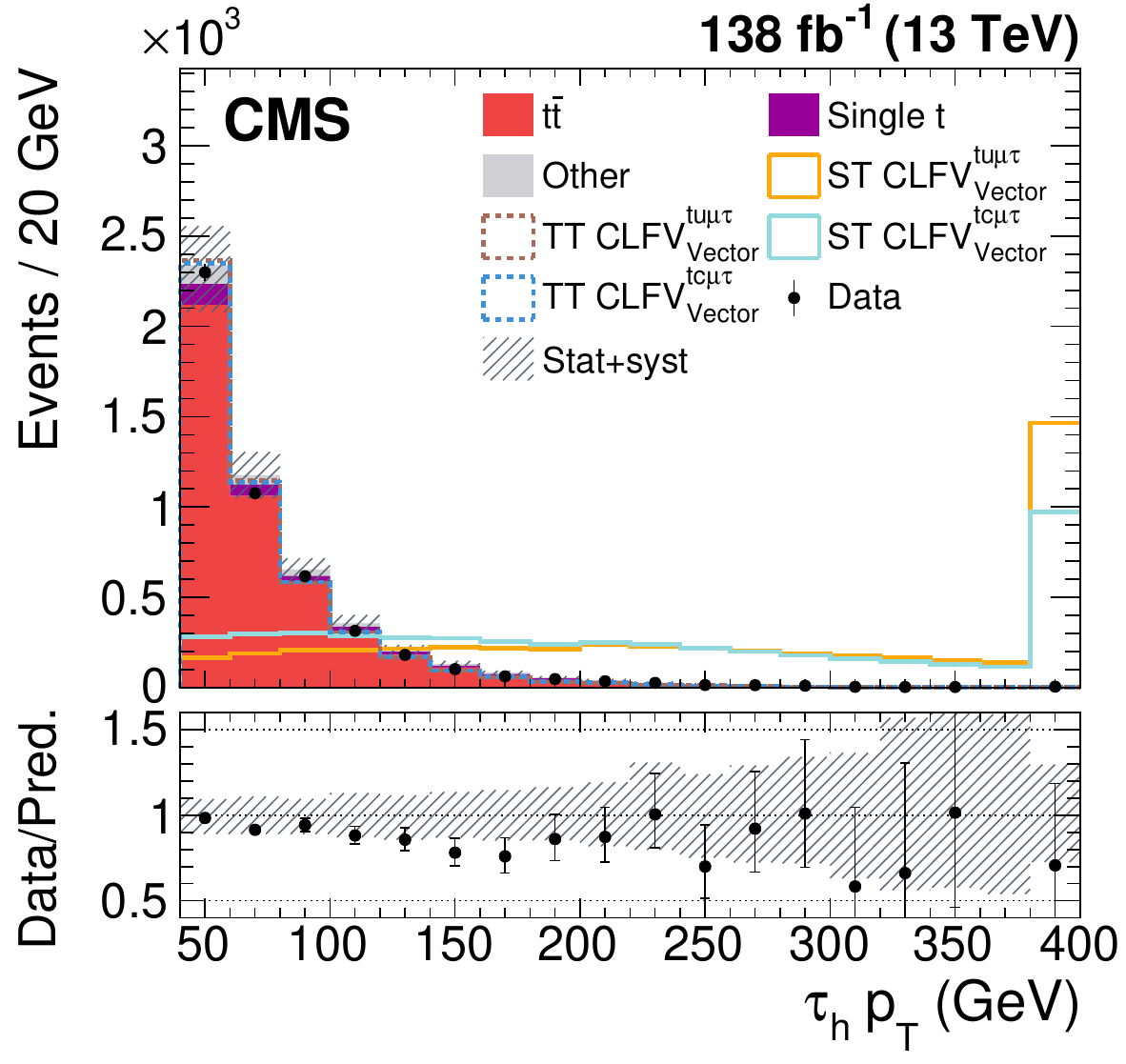} \\
\includegraphics[width=0.49\textwidth]{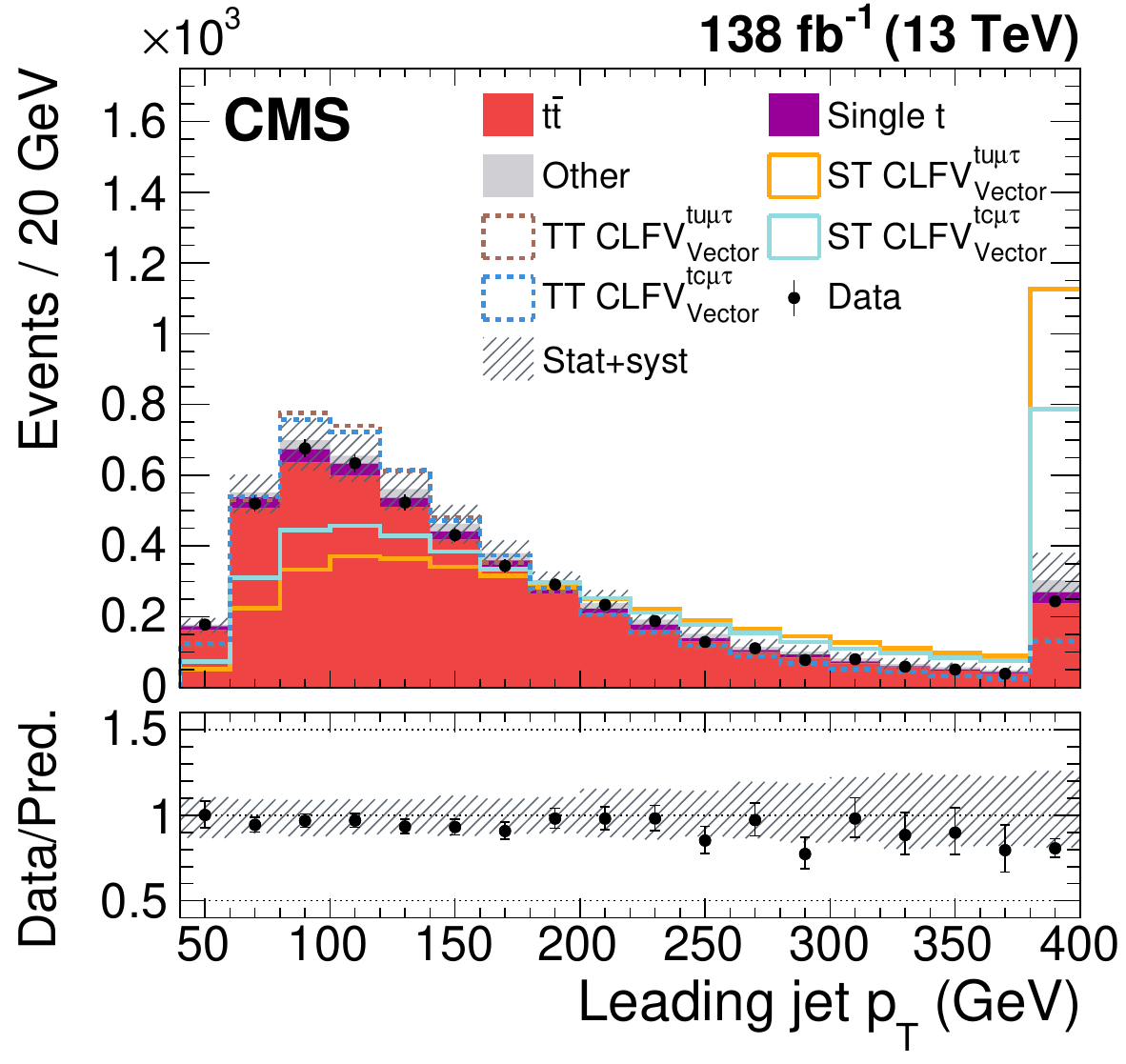}%
\hfill%
\includegraphics[width=0.49\textwidth]{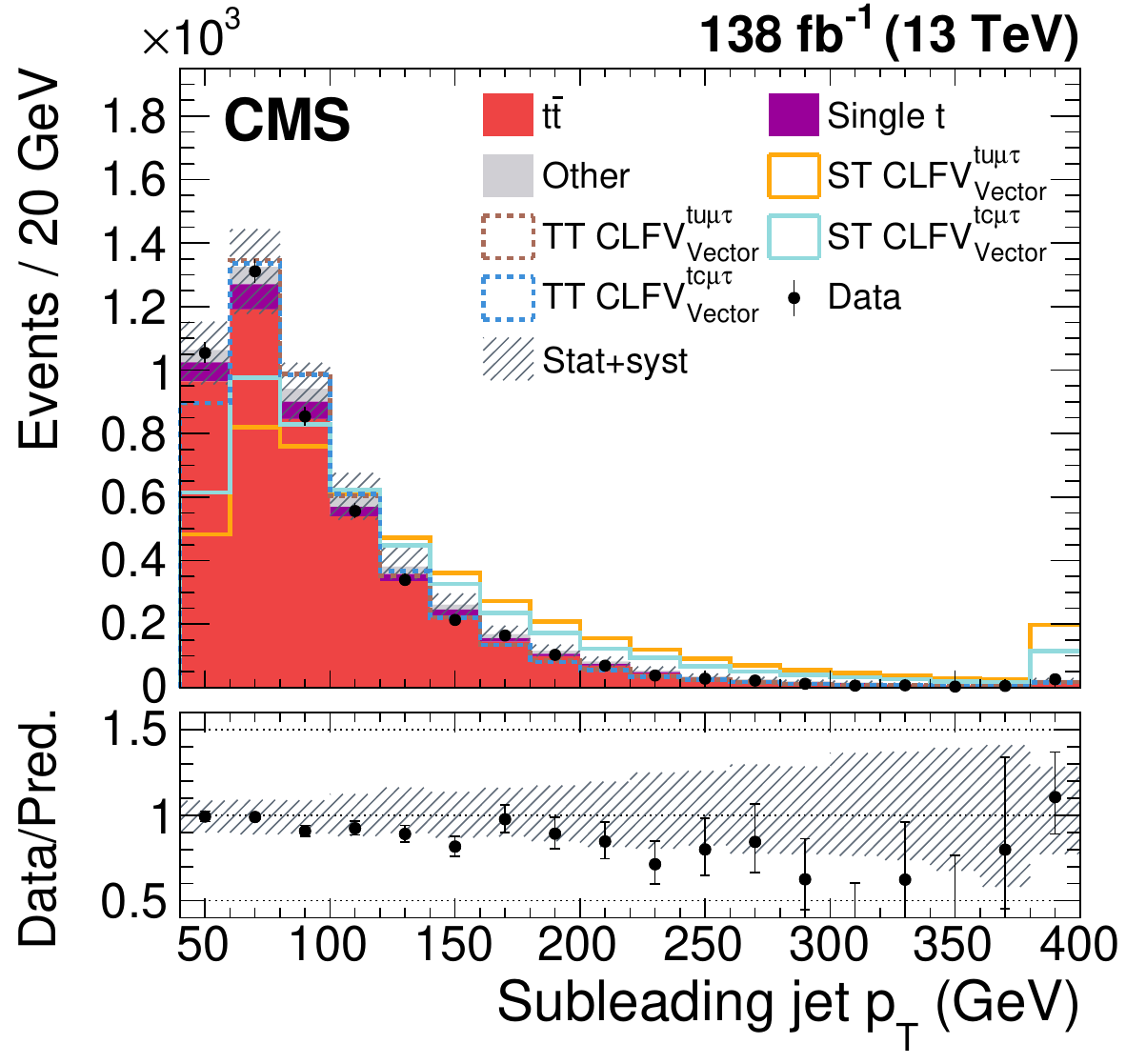}
\caption{%
    Distributions in \pt of the muon (upper left), \tauh (upper right),
    \pt-leading jet (lower left), and \pt-subleading jet (lower right) after all selection steps.
    The solid and dashed lines show the signal distributions for the vector Lorentz structure as an example, individually for each type of operator and interaction.
    The signals are normalized to the total number of events in data for visibility.
    The last bin in each histogram contains the overflow.
    The color filled histograms show the stacked background contributions.
    The data are shown as filled points.
    The shaded band displays the total uncertainty in the predicted SM background,
    consisting of statistical and systematic uncertainties added in quadrature.
    The panels below the distributions show the ratio of data to the background prediction.
}
\label{fig:datamc1_FF}
\end{figure}

We show distributions of important event properties in Fig.~\ref{fig:datamc1_FF},
after the event selection described in Section~\ref{sec:objects}.
Distributions obtained from data are compared to distributions from the sum of all SM background processes,
including the SFs from the ABCD method.
The signal processes are normalized to the total number of events observed in data for visibility of histograms.

The total number of expected events for CLFV signal and background processes,
where the background yields include \tauh misidentification SFs from the ABCD method,
are presented in Table~\ref{tab:yeildafterbkg} and compared to data.

\begin{table}[!tbh]
\centering
\topcaption{%
    Estimated event yields including the background corrections from the ABCD method
    discussed in Section~\ref{sec:bkgestimation}.
    The signal yields are computed assuming $C_a/\Lambda^2=1\TeV^{-2}$.
    The numbers shown correspond to observed events before the maximum likelihood fit
    described in Section~\ref{sec:results}.
    Only statistical uncertainties are shown, related to the size of the data sets.
}
\label{tab:yeildafterbkg}
\renewcommand{\arraystretch}{1.3}
\begin{tabular}{llr@{\,$\pm$\,}l}
	\multicolumn{2}{l}{Process} & \multicolumn{2}{c}{Event yield}\\
\hline
    \multirow{3}{*}{\STCLFV \tumutau}
	& Scalar & 535 & 1  \\
	& Vector & 2327 & 3  \\
	& Tensor & 9909 & 13  \\
    \multirow{3}{*}{\STCLFV \tcmutau}
	& Scalar & 32 & $<$1  \\
	& Vector & 129 & $<$1  \\
	& Tensor & 701 & 1  \\
    \multirow{3}{*}{\TTCLFV \tumutau}
	& Scalar & 1.1 & $<$0.1  \\
	& Vector & 8.2 & $<$0.1  \\
	& Tensor & 48 & $<$1  \\
    \multirow{3}{*}{\TTCLFV \tcmutau}
	& Scalar & 1.1 & $<$0.1  \\
	& Vector & 7.9 & $<$0.1  \\
	& Tensor & 45 & $<$1  \\[\cmsTabSkip]
	\ttbar && 4573 & 13  \\
	Single \PQt && 306 & 9  \\
	Other && 258 & 5  \\
	Total && 5136 & 17  \\[\cmsTabSkip]
	Data && \multicolumn{1}{r@{\,\phantom{$\pm$}\,}}{4810} &  \\
\end{tabular}
\end{table}

\section{Discrimination of signal and background}
\label{sec:signalextraction}

\subsection{Reconstruction of top quark and \texorpdfstring{\PW}{W} boson\label{sec:recotopW}}

In both \STCLFV and \TTCLFV signal events, a top quark is produced that decays hadronically
through the decay chain $\PQt\to\PQb\PW\to\PQb\PQq\PAQqpr$.
In contrast, background events mostly originate from \ttbar production. After applying the scale factors derived using the background estimation method described in Section~\ref{sec:bkgestimation}, 79.7\% of the background events arise from dileptonic decay channel, and 9.3\% from the lepton+jets decay channel.

{\tolerance=800
To benefit from this difference, the hadronic top quark decay is reconstructed using a
\chisq method.
The \chisq variable is constructed as
\begin{equation}
\chisq = \left(\frac{\mtop-\mbjj}{\sigma_{\PQt}}\right)^2 +
         \left(\frac{\mW-\mjj}{\sigma_{\PW}}\right)^2,
\label{eq:cat1chi2}
\end{equation}}
with the expected mass values $\mtop=173.95\GeV$ and $\mW=84.2\GeV$,
and resolutions $\sigma_{\PQt}=17.07\GeV$ and $\sigma_{\PW}=9.91\GeV$.
The mass and resolution values are computed in simulated \ttbar events
from detector-level jets matched to generator-level jets
from the top quark and \PW boson decays.
The reconstructed \PW boson candidate is formed by two jets with mass \mjj
and the selected \PQb-tagged jet is assigned to the reconstructed top quark candidate
with mass \mbjj.
The \chisq is calculated for all possible combinations of non-\PQb-tagged jets in an event.
For each event, the reconstruction hypothesis is chosen that results in the smallest
value of \chisq.

\begin{figure}[!tbp]
\centering
\includegraphics[width=0.49\textwidth]{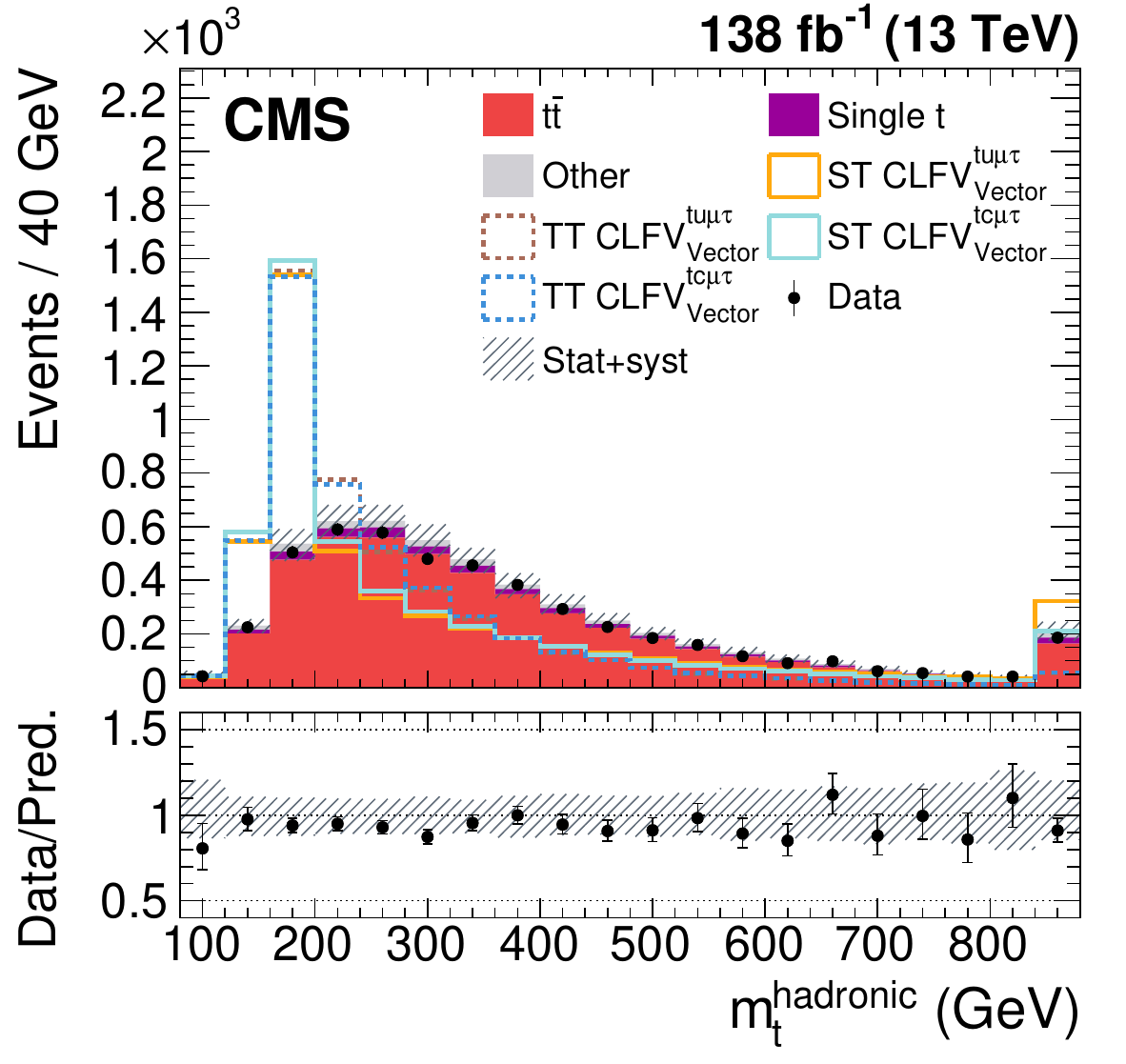}%
\hfill%
\includegraphics[width=0.49\textwidth]{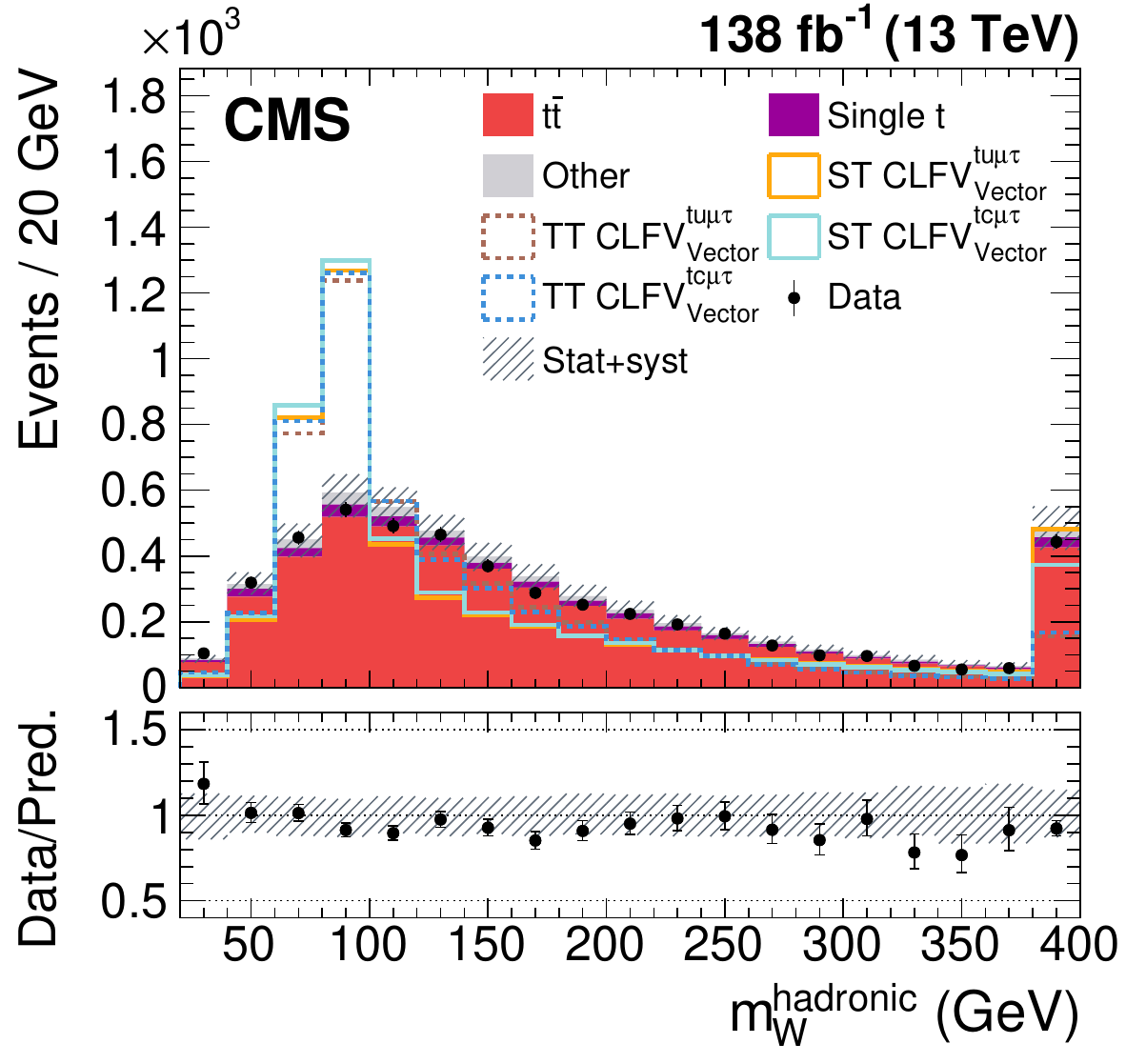} \\
\includegraphics[width=0.49\textwidth]{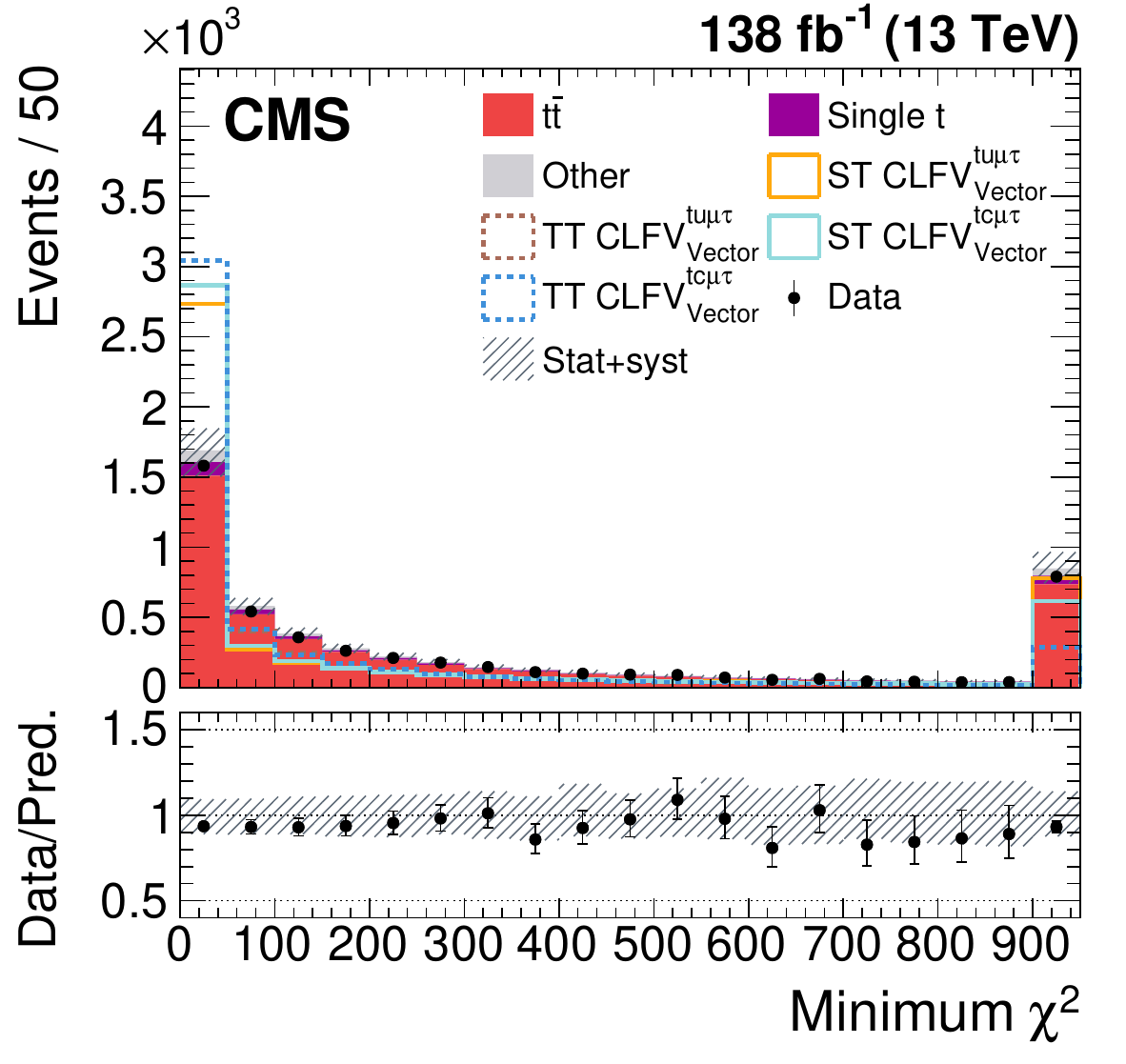}
\caption{%
    Distributions in the reconstructed top quark mass (upper left), \PW boson mass (upper right),
    and minimum \chisq (lower) from the top quark reconstruction.
    The solid and dashed lines show the signal distributions for the vector Lorentz structure as an example, individually for each type of operator and interaction.
    The signals are normalized to the total number of events in data for visibility.
    The color filled histograms show the stacked background contributions.
    The data are shown as filled points.
    The last bin in each histogram contains the overflow.
    The shaded band displays the total uncertainty in the predicted SM background,
    consisting of statistical and systematic uncertainties added in quadrature.
    The panels below the distributions show the ratio of data to the background prediction.
}
\label{fig:chi2_sttopreco2}
\end{figure}

The results of the reconstruction are presented in Fig.~\ref{fig:chi2_sttopreco2} for data, simulated signal and background events.
The SFs from the ABCD method are applied to the background simulation.
Distributions in signal events show more pronounced peaks around \mtop and \mW
in the reconstructed top quark and \PW boson masses, respectively, than the background distributions.
The latter have broader distributions because of large contributions from dileptonic \ttbar events.
Signal events show a noticeable peak at small values in the \chisq distribution, more pronounced
than the background. The data are described by the background simulation within the uncertainties.

\subsection{Multivariate analysis using a deep neural network}

The discrimination between CLFV signals and background processes is performed
using machine learning with a multiclass deep neural network (DNN) algorithm.

\begin{table}[!b]
\centering
\topcaption{%
    Input features of the DNN. The angular distance $\DR_{ij}$ between two objects $i$ and $j$ is defined as
	$\DR_{ij} = \sqrt{\smash[b]{(\Delta\eta_{ij})^2+(\Delta\phi_{ij})^2}}$, where $\Delta\eta_{ij}$ and $\Delta\phi_{ij}$ are the
    differences in pseudorapidity and azimuthal angle, respectively.
}
\label{tab:DNNinput}
\renewcommand{\arraystretch}{1.2}
\cmsTable{\begin{tabular}{lcl}
    Group & Variables & Description \\ \hline
    Muon (\PGm) & $p_{\mathrm{T}\PGm}$, $\eta_{\PGm}$ & \pt and $\eta$ of selected muon \\[\cmsTabSkip]
    Tau (\tauh) & $p_{\mathrm{T}\tauh}$, $\eta_{\tauh}$, $m_{\tauh}$ & \pt, $\eta$, and mass of selected \tauh \\[\cmsTabSkip]
    Muon+Tau (\mutauh) & $m_{\mutauh}$, $\Delta\eta_{\mutauh}$, $\Delta\phi_{\mutauh}$, $\DR_{\mutauh}$ & Mass and angular differences of \mutauh system \\[\cmsTabSkip]
    \multirow{4}{*}{Jets} & $p_{\mathrm{T}1}$, $p_{\mathrm{T}2}$, $p_{\mathrm{T}3}$ & \pt of jets ordered in decreasing \pt \\
    &$\eta_1$, $\eta_2$, $\eta_3$ & $\eta$ of jets ordered in decreasing \pt \\
    &$m_1$, $m_2$, $m_3$  & Mass of jets ordered in decreasing \pt \\
    &$b_1$, $b_2$, $b_3$ & \PQb tagging discriminant of jets ordered in decreasing \pt \\[\cmsTabSkip]
    Event & \ptmiss & Missing transverse momentum \\[\cmsTabSkip]
    Reconstruction of & \chisq, \mbjj, \mjj & Minimum \chisq and reconstructed \PQt and \PW masses\\
    \PQt and \PW & $\Delta\eta_{\jj}$, $\Delta\phi_{\jj}$, $\DR_{\jj}$ & Angular differences of jets used in \PW reconstruction \\
\end{tabular}}
\end{table}

The input features of the DNN comprise 28 variables, which include kinematic features of
individual objects and their combinations, and global event features such as
\ptmiss and variables from the \chisq reconstruction.
The input features together with a short description are given in Table~\ref{tab:DNNinput}.
The sensitivity of the DNN classifier is driven by the reconstructed top quark and \PW boson masses
\mbjj and \mjj, respectively, as well as the
mass of the \mutauh system $m_{\mutauh}$.
Following these, the \pt of the muon and \tauh candidates contribute to the classifier performance.

We optimize the DNN design and hyperparameters in terms of its accuracy score.
The model is implemented in \textsc{keras}~\cite{chollet2017keras} with the
\textsc{TensorFlow} backend~\cite{tensorflow2015-whitepaper}. The Adam~\cite{kingma2017adam} optimization algorithm
is used to minimize the sparse categorical cross-entropy loss function when updating the
DNN parameters during the training.
The DNN architecture consists of two hidden layers with 50 nodes each, where the weights are initialized
with randomly distributed weights following a normal distribution.
We use the ReLU~\cite{agarap2018deep} activation function and a batch size of 1024 events
in the training.
The number of training iterations is automatically determined and the training is stopped
if there is no improvement in the validation loss after 30 epochs.

The training of the DNN is performed using \STCLFV and \TTCLFV signal samples
and \ttbar background samples.
The DNN model is trained to predict the probability $p$ of an event belonging
to one of the three classes \STCLFV, \TTCLFV, and background.
Events after the final selection described in Section~\ref{sec:objects} are used for the training and the evaluation of the DNN.
The three probabilities are combined into a single variable
\begin{equation}
\text{DNN score} = \frac{0.1p(\TTCLFV)+0.9p(\STCLFV)}{p(\text{background})},
\label{Eq:dnnscore}
\end{equation}
which we found empirically to be an optimal choice for the best expected upper limits on the signal
cross sections.

A single DNN classifier is trained for all data-taking periods, equally combining the different
Lorentz structures and the two CLFV interactions \tumutau and \tcmutau.
We balance the number of signal and background events in the training, such that there
is no statistical bias from a dominant class.
About 150\,000 simulated events are passed to the DNN for each of the background and
the two signal classes, where 70\% are used for the training and 30\%
for the validation of the DNN.

\section{Systematic uncertainties}
\label{sec:systematic}

Systematic uncertainties are considered for experimental sources and the modelling of simulated samples,
including uncertainties in the predicted cross sections.
The experimental and modelling uncertainties generally change the
normalization and the shape of the distributions in the DNN score.
The different uncertainty sources are treated as nuisance parameters in a profile likelihood fit
in the statistical evaluation of the results.
The statistical uncertainties from the limited size of the simulated samples are treated
following the Barlow--Beeston-light approach~\cite{Conway:2011in}.

\subsection{Experimental uncertainties}

\paragraph*{Integrated luminosity}
The normalization of the signal and background predictions are varied within the uncertainties associated with the luminosity measurement in each year, which are 1.2~\cite{lumi2016}, 2.3~\cite{lumi2017}, and 2.5\%~\cite{lumi2018}
for 2016, 2017, and 2018, respectively. The total uncertainty for all years combined is 1.6\%,
taking into account correlations in the luminosity measurement between the individual data-taking periods.

\paragraph*{Pileup reweighting}
The uncertainty in the number of pileup interactions is estimated by varying the predicted
inelastic cross section of 69.2\unit{mb} by $\pm$4.6\%~\cite{CMS:2018mlc}.
This source is considered to be correlated across all data-taking periods.

\paragraph*{Trigger efficiency}
Data-to-simulation SFs for the trigger efficiency are varied within their uncertainties
for each data-taking period and are considered correlated between the periods.
In addition, a gradual timing shift of the ECAL was not properly propagated to the first-level trigger inputs
and caused an inefficiency by associating the input to the previous bunch crossing in 2016 and 2017,
known as ``prefiring''~\cite{CMS:2020cmk}.
A similar effect is present in the muon system due to the limited time resolution of the subdetectors,
mainly affecting the 2016 data, but also present in 2017 and 2018 where the effect is smaller.
Corresponding corrections are applied, and the uncertainties are evaluated by varying these corrections by $\pm$20\%~\cite{CMS:2024qjo}.
This uncertainty is uncorrelated between data-taking periods.

\paragraph*{Muon scale factors}
We apply corrections because of differences between data and simulation in the track reconstruction,
muon identification and isolation efficiencies. The corresponding uncertainties are derived
in the determination of these SFs in data samples enriched in $\PZ\to\PGm\PGm$ events~\cite{CMS:2018rym}.
In addition to the uncertainties in the SFs, as the reconstruction of high-\pt muons is more challenging and not fully covered by the standard muon identification uncertainties, an uncertainty for muons with \pt larger
than 200\GeV is implemented, which increases linearly with increasing \pt~\cite{CMS:2023iul}.
The uncertainties related to the muon identification and isolation are correlated across the data-taking periods.

\paragraph*{\tauh identification and energy scale}
Uncertainties in the identification efficiencies of the \tauh leptons for the discrimination
against jets, muons, and electrons consist of several sources related to statistical and
systematic components in the scale factor measurement.
The uncertainties related to the discrimination of \tauh candidates against jets are further divided with
respect to the data-taking period, the decay mode of the \tauh, and the \pt range~\cite{CMS:2022prd}.
The uncertainty in the \tauh energy scale is evaluated by shifting the \tauh momentum in simulated events
within its uncertainty~\cite{CMS:2018jrd,CMS:2022prd}.
The uncertainties in the \tauh identification against jets are partly correlated across
the data-taking periods and the \tauh decay modes, where the latter are defined by the number of charged and neutral hadrons in the \tauh decay products, depending on the source of uncertainty.
Other uncertainties related to the \tauh identification and reconstruction are uncorrelated.

\paragraph*{Background estimation for jets misidentified as \texorpdfstring{\tauh}{tauh}}
The statistical uncertainty arising from the determination of the scaling factors for
events with jets misidentified as \tauh candidates is considered.
In addition, a closure test is performed using simulated events for individual decay modes
of the \tauh candidate and the different data-taking periods.
The difference between the number of simulated background events and the number of events
predicted by the ABCD method is considered as an uncertainty, and amounts to 2--3\%.
This uncertainty is uncorrelated for all data-taking periods.

\paragraph*{Jet energy scale and resolution}
The uncertainties in the JES are estimated by shifting the jet momenta in simulated events
within their uncertainties for several sources of the JES correction.
The JER uncertainty is evaluated by changing the resolution in the simulation by one
standard deviation~\cite{CMS:2016lmd}.
An inefficiency caused by a non-functional HCAL module in 2018 results in an additional
uncertainty~\cite{CMS:2021ime}, which is estimated by scaling down the jet energy in the affected detector region.
The JES and JER variations change the \pt and the energy of jets and are propagated to
\ptvecmiss.
The JES uncertainties are correlated for some sources, while the uncertainties in JER are
uncorrelated for all data-taking periods.

\paragraph*{\PQb tagging}
The \PQb tagging SFs are applied to correct the tagging efficiency in simulation.
The uncertainties related to the correction of the \PQb tagging discriminant are evaluated separately for light-quark and gluon jets, and for \PQb and \PQc jets~\cite{BTV-16-002}.
In addition, for each variation of the JES, dedicated scale factors are calculated and applied.
The \PQb tagging uncertainty sources include effects from the flavour of jets
and statistical uncertainties that arise from the scale factor measurement.
The statistical components are uncorrelated between data-taking periods,
while the others are fully correlated.

\subsection{Modelling uncertainties}
The modelling uncertainties are fully correlated through all data-taking periods.

\paragraph*{PDF and the strong coupling}
Uncertainties in the PDF are estimated by reweighting events with the eigenvector variations
of the NNPDF~3.1 set~\cite{NNPDF:2017mvq}.
The PDF uncertainty is included for the \ttbar background and signal samples.
In addition, an uncertainty from the choice of the strong coupling \alpS in the PDF
set is taken into account~\cite{Butterworth:2015oua} for the \ttbar background simulation.

\paragraph*{Missing higher orders}
The uncertainty from missing higher orders in the calculation of matrix elements (ME)
for signal and \ttbar background processes is estimated by varying the renormalization (\muR)
and factorization (\muF) scales.
We consider variations of the \muR and \muF by factors of 0.5 and 2. Specifically, we vary \muR while leaving \muF unchanged, and vice versa, as well as performing simultaneous variations of both scales.
The \pt of the muon and \tauh candidates is divided into two ranges with 50--100\,\GeV and $>$100\,\GeV for each, resulting in four distinct combinations considered in the uncertainty evaluation, to address expected differences arising from additional
jets in the \TTCLFV signals at high lepton \pt.

\paragraph*{Parton shower}
Uncertainties in the initial- and final-state radiation (ISR and FSR) are estimated by
varying the shower scales in the parton shower (PS) by factors of 2 and 0.5 separately for the ISR and FSR.
This uncertainty is included for signal and \ttbar background samples.

\paragraph*{Cross sections}
The uncertainties in the predicted cross sections are treated as several different normalization uncertainties.
In \ttbar production, scale variations and uncertainties in the PDF amount to a total
uncertainty of 4.4\%~\cite{Czakon:2011xx}.
An uncertainty of 10\% is assigned to single top quark production and to all other background processes.

\paragraph*{ME--PS matching}
The damping of high-\pt radiation in the matching of the ME and PS simulation
is regulated by the parameter $h_{\text{damp}}$ in \POWHEG.
We estimate the uncertainty from the ME-PS matching by shifting
$h_{\text{damp}} = 1.38\,^{+0.93}_{-0.51}\,\mtop$ within its uncertainties~\cite{Sirunyan:2019dfx}, where
$\mtop=172.5\GeV$.
This uncertainty affects the \ttbar background.

\paragraph*{Underlying event}
The uncertainty related to the CP5 underlying event tune~\cite{Sirunyan:2019dfx} of \PYTHIA
is evaluated from \ttbar samples generated with modified tune parameters.

\paragraph*{Top quark \pt reweighting}
The uncertainty arising from the modelling of the top quark \pt spectrum
is evaluated by changes in the \muR and \muF scales in the NNLO
calculation~\cite{Czakon:2017wor} for the determination of the correction.

\section{Results}
\label{sec:results}

Distributions in the DNN score are investigated in the search for CLFV interactions.
We set upper limits at 95\% \CL on the coupling modifier
$\hat{\mu}_a$, which translates to limits on the squared Wilson coefficients $(C_a/\Lambda^2)^2$.
To set these limits, we use a modified frequentist \CLs method~\cite{Junk:1999kv,Read2002},
which uses the profile likelihood ratio as the test statistic.
The distributions in this test statistic are determined using an asymptotic approximation~\cite{Cowan:2010js}.
A binned maximum likelihood fit is constructed using the measured distribution in the DNN score,
the simulated background processes, and the predicted CLFV signals.
The systematic uncertainties described in Section~\ref{sec:systematic} are included as nuisance parameters.
The statistical analysis is performed using the CMS statistical analysis tool \textsc{Combine}~\cite{CMS:2024onh},
which is based on the \textsc{RooFit}~\cite{Verkerke:2003ir} and \textsc{RooStats}~\cite{Moneta:2010pm} frameworks.
Upper limits on the individual coupling modifiers are calculated by performing a
simultaneous fit to all data-taking periods.

\begin{figure}[!t]
\centering
\includegraphics[width=0.49\textwidth]{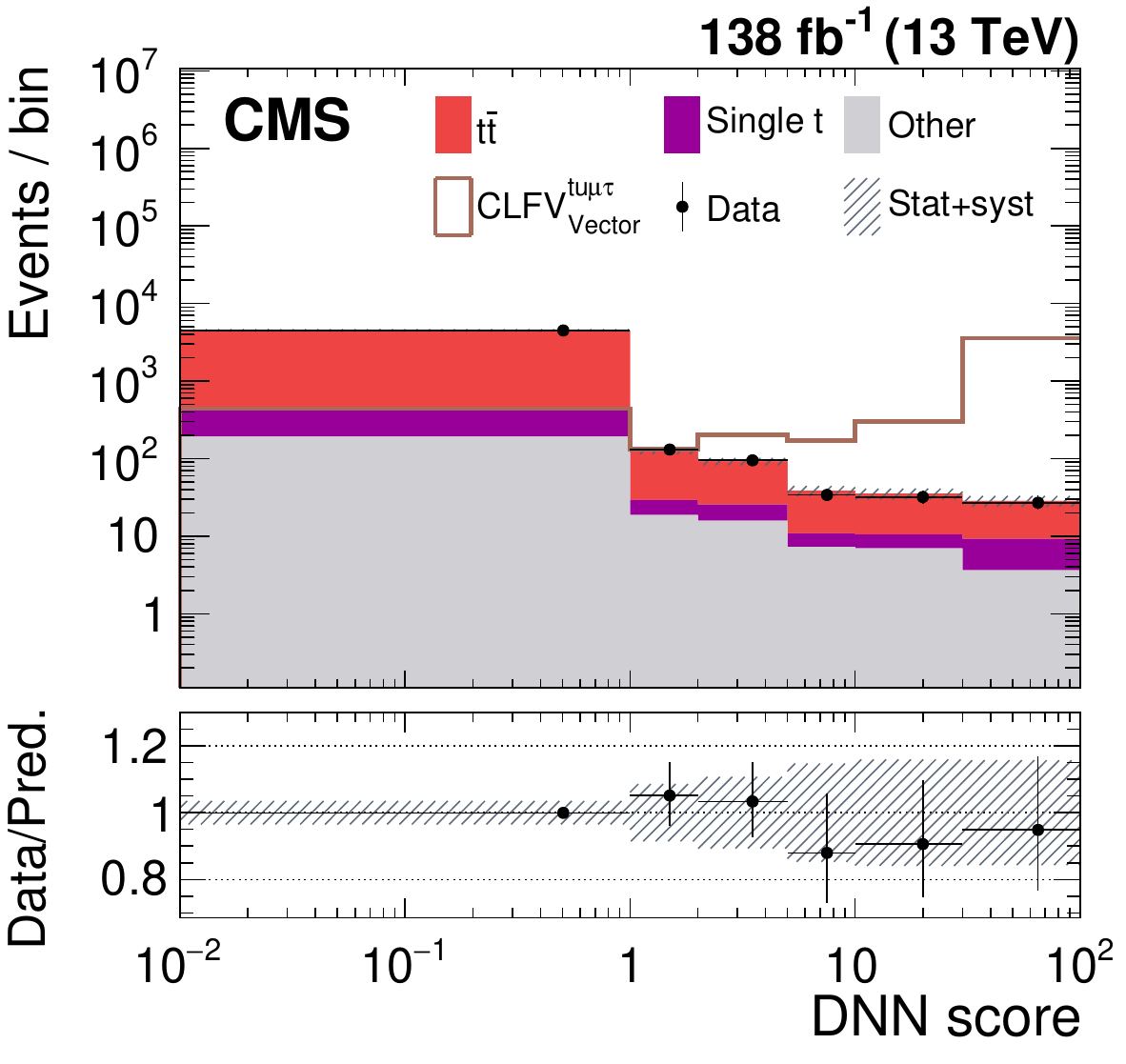}%
\hfill%
\includegraphics[width=0.49\textwidth]{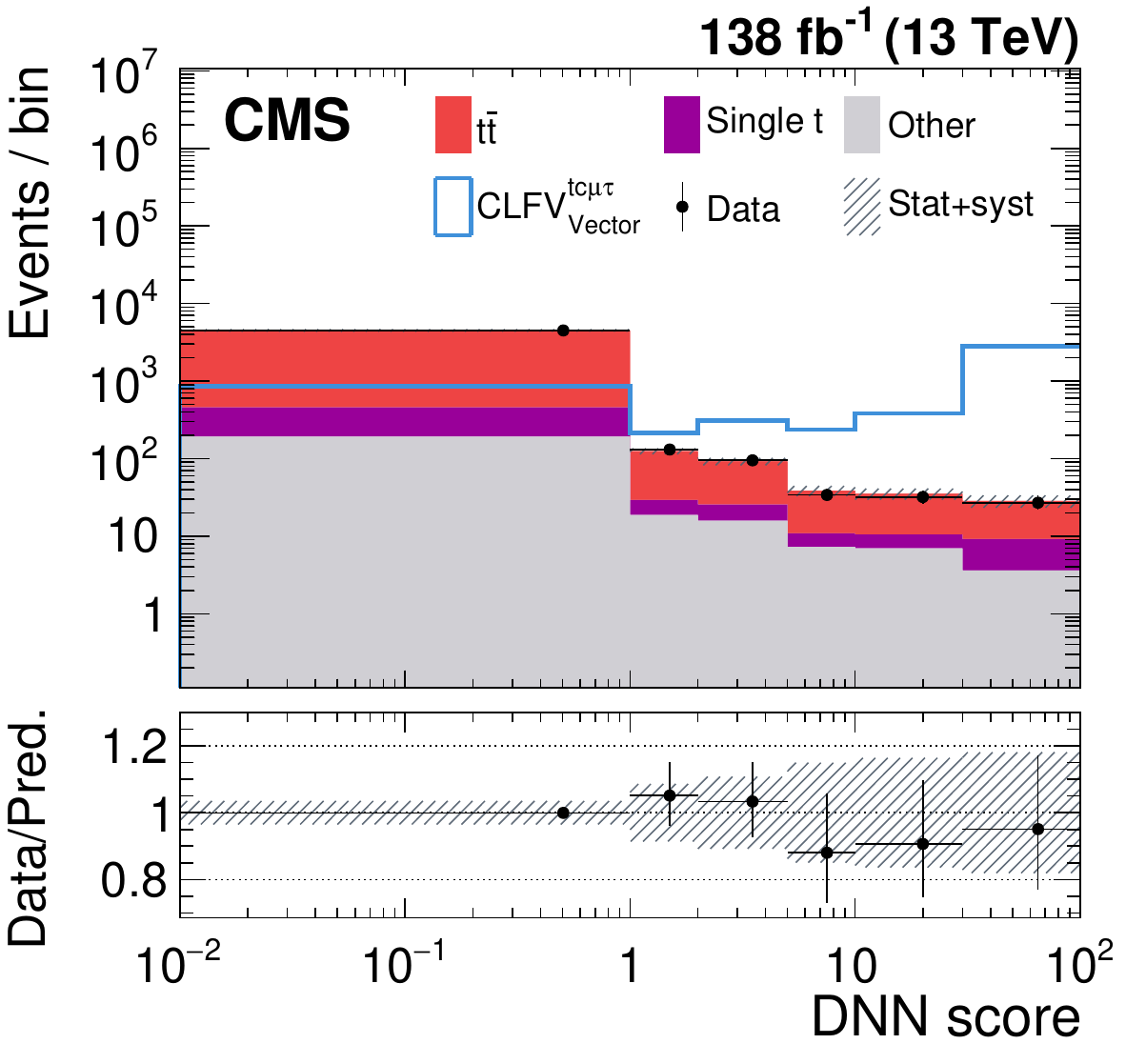}
\caption{%
    Combined distributions in the DNN score after the profile likelihood fit for all data-taking periods
    for the vector operators with \tumutau (left) and \tcmutau (right) couplings.
    The signal distributions are normalized to the total number of events observed in the data.
    The first and last bins of each histogram contain the underflow and overflow, respectively.
    The hatched bands represent the total post-fit uncertainties in the background predictions, including statistical and systematic sources.
    The panels below the distributions show the ratio of data to the background prediction.
}
\label{fig:DNNoutput_postFit_p_st_tt_ob_Run2}
\end{figure}

Distributions in the DNN score are shown for two signal hypotheses in Fig.~\ref{fig:DNNoutput_postFit_p_st_tt_ob_Run2}
after the binned maximum likelihood fit for all data-taking periods combined.
At the highest value of the DNN score, corresponding to the histogram bin with the highest
sensitivity to CLFV interactions, uncertainties of 15--20\% in the background predictions
are obtained, comparable to the statistical uncertainties in data.
We observe no significant deviation from the SM predictions.

We derive one-dimensional upper limits on individual Wilson coefficients.
The limits are calculated by taking the square root of the upper limits on the corresponding
$\hat{\mu}_a$ for each CLFV interaction operator with $\Lambda=1\TeV$.
Upper limits on the branching fractions $\BR(\tTomutauq)$, with $\PQq=\PQu$ or \PQc, are calculated
using Eq.~(2) from Ref.~\cite{Kile:2008rp}.
Signal processes are considered separately for each Lorentz structure and quark flavour,
assuming only one non-vanishing coupling at a time.
Limits on the CLFV signal cross sections $\sigma$ are calculated by multiplying the limit on the coupling
modifier with the sum of the \STCLFV and \TTCLFV cross sections from Table~\ref{tab:totcalc_xsec}.
In Table~\ref{tab:2limit}, the observed and expected upper limits on cross sections,
Wilson coefficients, and branching fractions for each interaction are listed with the
central probability intervals containing 68\% of the expected upper limits.

\begin{table}[!tbp]
\centering
\topcaption{%
    The 95\% \CL observed and expected upper limits on CLFV cross sections,
    Wilson coefficients $C_{\tqmutau}$, and branching fractions for different types
    of interactions and Lorentz structures. The expected upper limits are shown in
    parentheses after the observed limits. The central probability intervals containing
    68\% of the expected upper limits are given in square brackets below the upper limits.
}
\label{tab:2limit}
\renewcommand{\arraystretch}{1.2}
\begin{tabular}{ccccc}
    Interaction & Type & $\sigma$ [fb] & $C_{\tqmutau}/\Lambda^2$ [$\TeVns^{-2}$] & $\mathcal{B}(\tTomutauq)$ [$10^{-6}$] \\[2pt] \hline
    \multirow{6}{*}{\tumutau}
        & \multirow{2}{*}{Scalar} & 2.04 (2.34) & 0.18 (0.19) & 0.04 (0.05) \\
        & & [1.57, 3.59] & [0.16, 0.24] & [0.03, 0.07] \\[0.3\cmsTabSkip]
        & \multirow{2}{*}{Vector} & 2.38 (2.75) & 0.09 (0.10) & 0.08 (0.09) \\
        & & [1.86, 4.21] & [0.08, 0.12] & [0.06, 0.14] \\[0.3\cmsTabSkip]
        & \multirow{2}{*}{Tensor} & 2.83 (3.33) & 0.05 (0.05) & 0.19 (0.14) \\
        & & [2.26, 5.06] & [0.04, 0.06] & [0.09, 0.21] \\[\cmsTabSkip]
    \multirow{6}{*}{\tcmutau}
        & \multirow{2}{*}{Scalar} & 4.27 (5.02) & 0.82 (0.89) & 0.81 (0.95) \\
        & & [3.29, 8.14] & [0.72, 1.13] & [0.63, 1.55] \\[0.3\cmsTabSkip]
        & \multirow{2}{*}{Vector} & 7.21 (8.55) & 0.42 (0.46) & 1.71 (2.03) \\
        & & [5.66, 13.73] & [0.37, 0.58] & [1.34, 3.26] \\[0.3\cmsTabSkip]
        & \multirow{2}{*}{Tensor} & 7.93 (9.63) & 0.19 (0.21) & 2.05 (2.49) \\
        & & [6.43, 15.20] & [0.17, 0.26] & [1.66, 3.94] \\
\end{tabular}
\end{table}

\begin{figure}[!tbp]
\centering
\includegraphics[width=0.49\textwidth]{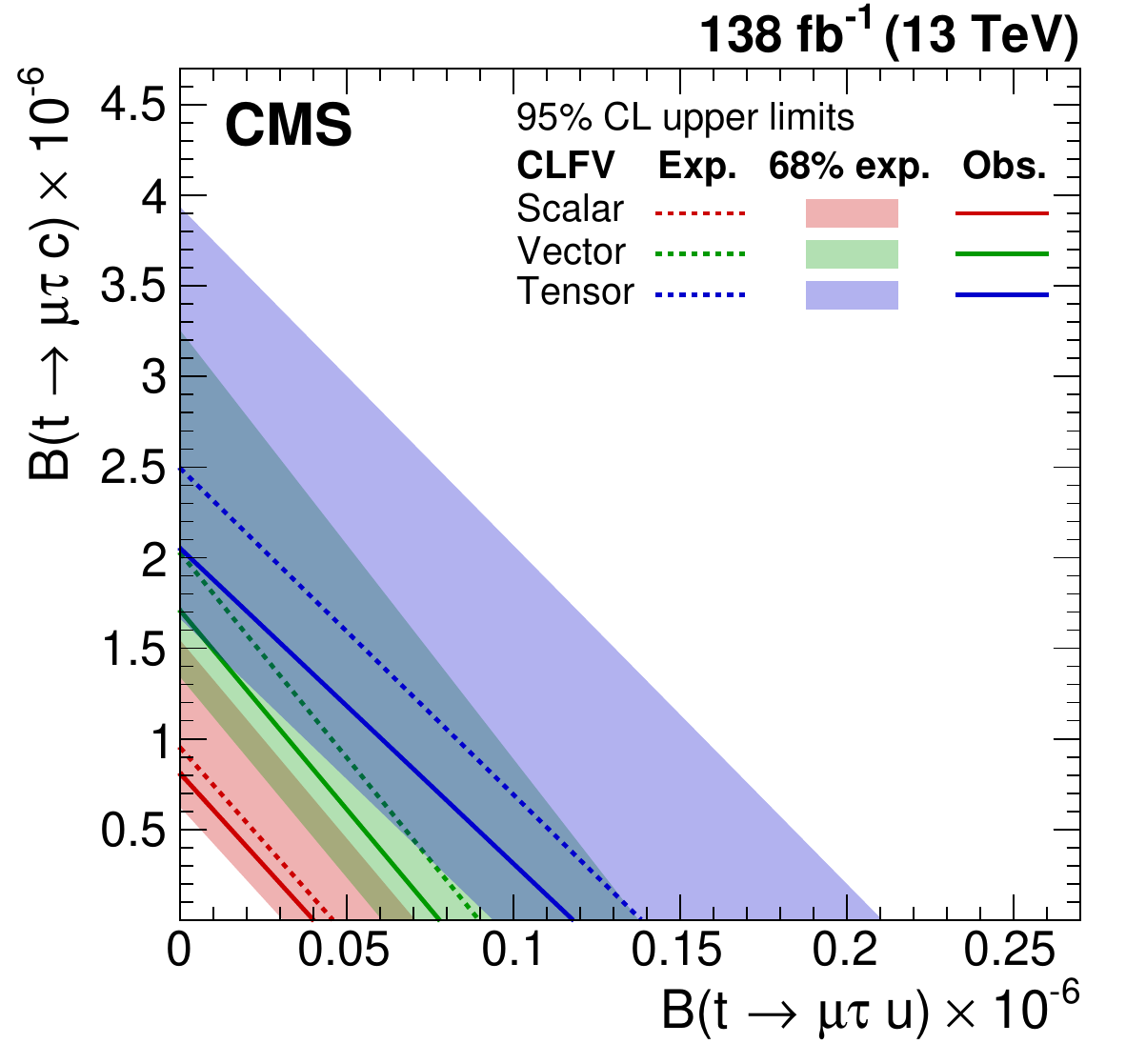}%
\hfill%
\includegraphics[width=0.49\textwidth]{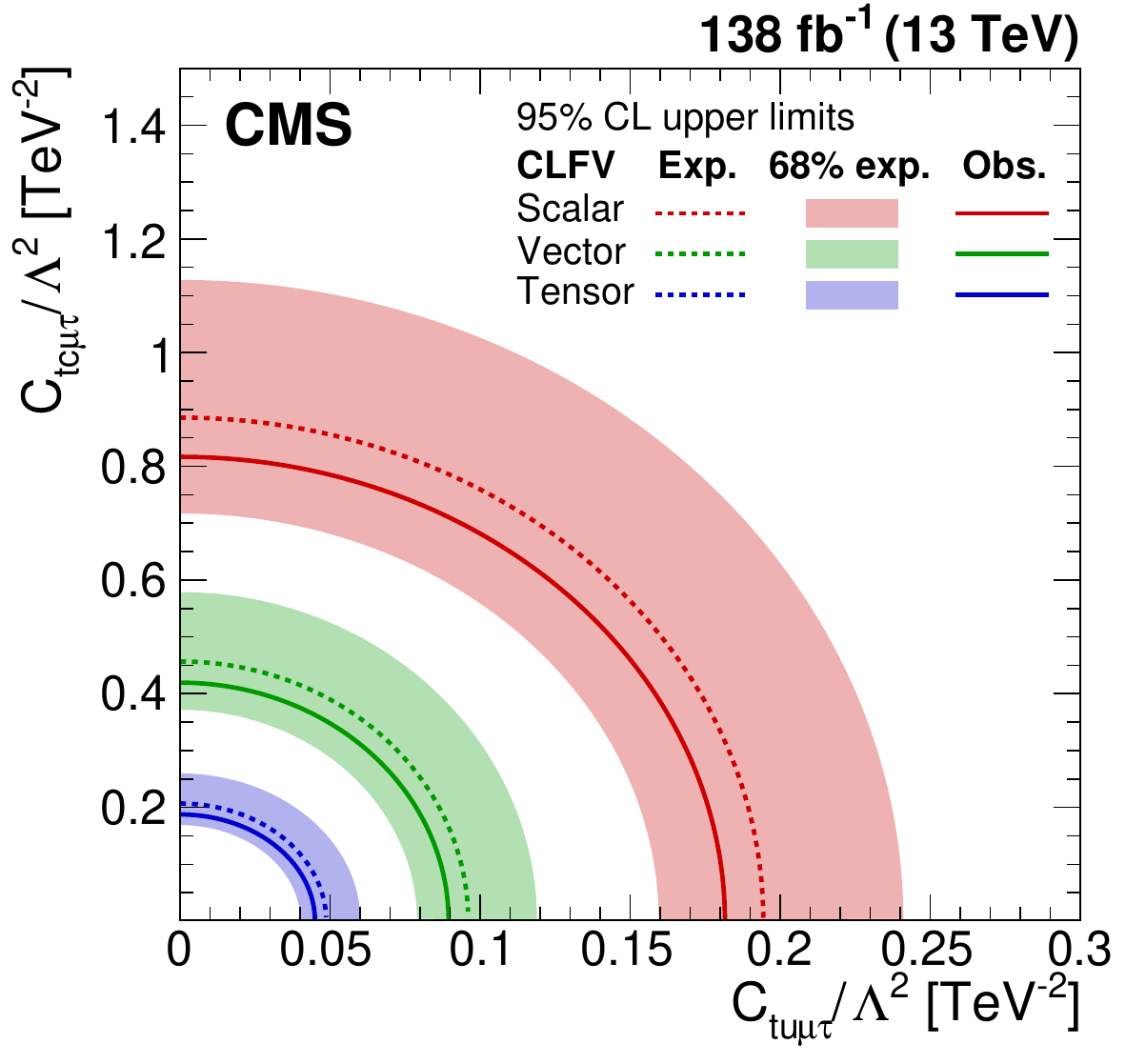}
\caption{
    Exclusion contours for the observed and expected 95\% \CL upper limits and central probability
    intervals containing 68\% of the expected upper limits for the branching fractions (left)
    and Wilson coefficients (right) corresponding to the \tumutau and \tcmutau
    couplings for scalar, vector and tensor Lorentz structures.
}
\label{fig:2Dlimit}
\end{figure}

In Fig.~\ref{fig:2Dlimit}, we show exclusion contours calculated from the results of Table~\ref{tab:2limit}.
The contours are obtained with an interpolation by assuming a linear relationship between
the limits on $\BR(\tTomutauc)$ and $\BR(\tTomutauu)$ for each operator.
The contours are shown in the plane of branching fractions (left) and Wilson coefficients (right)
for the \tumutau and \tcmutau couplings for scalar, vector, and tensor operators.
In the \STCLFV processes, the cross sections corresponding to the \tumutau coupling
are larger than those for the \tcmutau coupling, resulting in more stringent
constraints on the corresponding Wilson coefficients.
The largest impact on the sensitivity of the results originates from the limited sample
size in the last bin of the DNN score and the ME scale variations.

\section{Summary}
\label{sec:conclusion}

A search for charged-lepton flavour violation (CLFV) in the top quark sector has been presented.
The search uses data corresponding to an integrated luminosity of 138\fbinv collected by the CMS experiment
during 2016--2018 in proton-proton (\pp) collisions at $\sqrt{s}=13\TeV$.

Interactions of a top quark with a muon, a \PGt lepton, and an up-type quark \PQu or \PQc
are considered, where the scale of new physics responsible for CLFV is assumed to be
larger than the energy of \pp collisions at the LHC.
The signal extraction is performed using measured distributions in a multiclass discriminator
obtained with a deep neural network.
No significant deviation is observed from the standard model background prediction.
Upper limits on the signal cross sections are set at 95\% confidence level (\CL).
The limits are interpreted in terms of CLFV branching fractions (\BR) of the top quark,
resulting in $\BR(\tTomutauu)<(0.04$, 0.08, and $0.12)\times10^{-6}$,
and $\BR(\tTomutauc)<(0.81$, 1.71, and $2.05)\times10^{-6}$ at 95\% \CL
for scalar, vector, and tensor-like operators, respectively.
This search complements previous CMS results involving $\Pe\PGm$ CLFV
interactions~\cite{TOP-19-006,CMS:2023iul}
and results in more stringent upper limits on Wilson coefficients in
an effective field theory by approximately a factor of two compared
to the latest experimental results involving $\PGm\PGt$ CLFV interactions~\cite{ATLAS:2024njy}.

\begin{acknowledgments}
We congratulate our colleagues in the CERN accelerator departments for the excellent performance of the LHC and thank the technical and administrative staffs at CERN and at other CMS institutes for their contributions to the success of the CMS effort. In addition, we gratefully acknowledge the computing centres and personnel of the Worldwide LHC Computing Grid and other centres for delivering so effectively the computing infrastructure essential to our analyses. Finally, we acknowledge the enduring support for the construction and operation of the LHC, the CMS detector, and the supporting computing infrastructure provided by the following funding agencies: SC (Armenia), BMBWF and FWF (Austria); FNRS and FWO (Belgium); CNPq, CAPES, FAPERJ, FAPERGS, and FAPESP (Brazil); MES and BNSF (Bulgaria); CERN; CAS, MoST, and NSFC (China); MINCIENCIAS (Colombia); MSES and CSF (Croatia); RIF (Cyprus); SENESCYT (Ecuador); ERC PRG, RVTT3 and MoER TK202 (Estonia); Academy of Finland, MEC, and HIP (Finland); CEA and CNRS/IN2P3 (France); SRNSF (Georgia); BMBF, DFG, and HGF (Germany); GSRI (Greece); NKFIH (Hungary); DAE and DST (India); IPM (Iran); SFI (Ireland); INFN (Italy); MSIP and NRF (Republic of Korea); MES (Latvia); LMTLT (Lithuania); MOE and UM (Malaysia); BUAP, CINVESTAV, CONACYT, LNS, SEP, and UASLP-FAI (Mexico); MOS (Montenegro); MBIE (New Zealand); PAEC (Pakistan); MES and NSC (Poland); FCT (Portugal);  MESTD (Serbia); MICIU/AEI and PCTI (Spain); MOSTR (Sri Lanka); Swiss Funding Agencies (Switzerland); MST (Taipei); MHESI and NSTDA (Thailand); TUBITAK and TENMAK (Turkey); NASU (Ukraine); STFC (United Kingdom); DOE and NSF (USA).

\hyphenation{Rachada-pisek} Individuals have received support from the Marie-Curie programme and the European Research Council and Horizon 2020 Grant, contract Nos.\ 675440, 724704, 752730, 758316, 765710, 824093, 101115353, 101002207, and COST Action CA16108 (European Union); the Leventis Foundation; the Alfred P.\ Sloan Foundation; the Alexander von Humboldt Foundation; the Science Committee, project no. 22rl-037 (Armenia); the Fonds pour la Formation \`a la Recherche dans l'Industrie et dans l'Agriculture (FRIA-Belgium); the Beijing Municipal Science \& Technology Commission, No. Z191100007219010 and Fundamental Research Funds for the Central Universities (China); the Ministry of Education, Youth and Sports (MEYS) of the Czech Republic; the Shota Rustaveli National Science Foundation, grant FR-22-985 (Georgia); the Deutsche Forschungsgemeinschaft (DFG), among others, under Germany's Excellence Strategy -- EXC 2121 ``Quantum Universe" -- 390833306, and under project number 400140256 - GRK2497; the Hellenic Foundation for Research and Innovation (HFRI), Project Number 2288 (Greece); the Hungarian Academy of Sciences, the New National Excellence Program - \'UNKP, the NKFIH research grants K 131991, K 133046, K 138136, K 143460, K 143477, K 146913, K 146914, K 147048, 2020-2.2.1-ED-2021-00181, TKP2021-NKTA-64, and 2021-4.1.2-NEMZ\_KI-2024-00036 (Hungary); the Council of Science and Industrial Research, India; ICSC -- National Research Centre for High Performance Computing, Big Data and Quantum Computing and FAIR -- Future Artificial Intelligence Research, funded by the NextGenerationEU program (Italy); the Latvian Council of Science; the Ministry of Education and Science, project no. 2022/WK/14, and the National Science Center, contracts Opus 2021/41/B/ST2/01369 and 2021/43/B/ST2/01552 (Poland); the Funda\c{c}\~ao para a Ci\^encia e a Tecnologia, grant CEECIND/01334/2018 (Portugal); the National Priorities Research Program by Qatar National Research Fund; MICIU/AEI/10.13039/501100011033, ERDF/EU, "European Union NextGenerationEU/PRTR", and Programa Severo Ochoa del Principado de Asturias (Spain); the Chulalongkorn Academic into Its 2nd Century Project Advancement Project, and the National Science, Research and Innovation Fund via the Program Management Unit for Human Resources \& Institutional Development, Research and Innovation, grant B39G670016 (Thailand); the Kavli Foundation; the Nvidia Corporation; the SuperMicro Corporation; the Welch Foundation, contract C-1845; and the Weston Havens Foundation (USA).
\end{acknowledgments}

\bibliography{auto_generated}
\cleardoublepage \appendix\section{The CMS Collaboration \label{app:collab}}\begin{sloppypar}\hyphenpenalty=5000\widowpenalty=500\clubpenalty=5000
\cmsinstitute{Yerevan Physics Institute, Yerevan, Armenia}
{\tolerance=6000
A.~Hayrapetyan, A.~Tumasyan\cmsAuthorMark{1}\cmsorcid{0009-0000-0684-6742}
\par}
\cmsinstitute{Institut f\"{u}r Hochenergiephysik, Vienna, Austria}
{\tolerance=6000
W.~Adam\cmsorcid{0000-0001-9099-4341}, J.W.~Andrejkovic, L.~Benato\cmsorcid{0000-0001-5135-7489}, T.~Bergauer\cmsorcid{0000-0002-5786-0293}, S.~Chatterjee\cmsorcid{0000-0003-2660-0349}, K.~Damanakis\cmsorcid{0000-0001-5389-2872}, M.~Dragicevic\cmsorcid{0000-0003-1967-6783}, P.S.~Hussain\cmsorcid{0000-0002-4825-5278}, M.~Jeitler\cmsAuthorMark{2}\cmsorcid{0000-0002-5141-9560}, N.~Krammer\cmsorcid{0000-0002-0548-0985}, A.~Li\cmsorcid{0000-0002-4547-116X}, D.~Liko\cmsorcid{0000-0002-3380-473X}, I.~Mikulec\cmsorcid{0000-0003-0385-2746}, J.~Schieck\cmsAuthorMark{2}\cmsorcid{0000-0002-1058-8093}, R.~Sch\"{o}fbeck\cmsAuthorMark{2}\cmsorcid{0000-0002-2332-8784}, D.~Schwarz\cmsorcid{0000-0002-3821-7331}, M.~Sonawane\cmsorcid{0000-0003-0510-7010}, W.~Waltenberger\cmsorcid{0000-0002-6215-7228}, C.-E.~Wulz\cmsAuthorMark{2}\cmsorcid{0000-0001-9226-5812}
% ...
\par}
\vskip\cmsinstskip
\dag:~Deceased\\
$^{1}$Also at Yerevan State University, Yerevan, Armenia\\
$^{2}$Also at TU Wien, Vienna, Austria\\
% ...
\end{sloppypar}
\end{document}